\documentclass[12pt]{article}
\usepackage{latexsym,kec}
\usepackage{graphicx}
\begin{document}
\pagestyle{empty}
\begin{flushright}
{\small NMCPP/99-17
}\\
\end{flushright}
\begin{center}
{\bf\Large Neutrinos Are Nearly Dirac Fermions}\\
\vspace*{0.50cm}
{\bf Kevin Cahill\footnote{kevin@kevin.phys.unm.edu
\quad http://kevin.phys.unm.edu/\(\tilde{\ }\)kevin/}}\\
\vspace{.15cm}
{\small New Mexico Center for Particle Physics\\
Department of Physics and Astronomy\\
University of New Mexico,
Albuquerque, New Mexico 87131-1156}\\
\vspace*{0.05cm}
\end{center}
\begin{quote}
\begin{center}
{\bf Abstract}
\end{center}
Neutrino masses and mixings are analyzed in terms
of left-handed fields and a
\(6\times6\) complex symmetric mass matrix \(\cM\)
whose singular values are the neutrino masses.
An angle \( x_\nu\) 
characterizes the kind of the neutrinos,
with \( x_\nu=0\) for Dirac neutrinos
and \( x_\nu=\pi/2\) for Majorana neutrinos.
If \(  x_\nu = 0 \),
then baryon-minus-lepton number is conserved.
When \(  x_\nu \approx 0 \),
the six neutrino masses coalesce
into three nearly degenerate pairs.
Thus the smallness of the differences
in neutrino masses exhibited in the
solar and atmospheric neutrino experiments
and the stringent limits on neutrinoless double beta decay
are naturally explained if \(B-L\) is approximately conserved
and neutrinos are nearly Dirac fermions.
If one sets \( \sin x_\nu = 0.003\),
suppresses inter-generational mixing,
and imposes a quark-like mass hierarchy,
then one may fit the essential features of the
solar, reactor, and atmospheric neutrino data
with otherwise random mass matrices \(\cM\) in the eV range.
\par
This \(B-L\) model leads to these predictions:
neutrinos oscillate mainly between flavor eigenfields
and sterile eigenfields, and so
the probabilities of the appearance of neutrinos
or antineutrinos are very small;
neutrinos may well be of cosmological importance;
in principle the disappearance of \(\nu_\tau\)
should be observable;
and \(0\nu\b\b\) decay is suppressed by
an extra factor of \(10^{-5}\)
and hence will not be seen in the
Heidelberg/Moscow, IGEX, GENIUS, or CUORE
experiments.
\end{quote}
\vfill
\vfill\eject

\setcounter{equation}{0}
\pagestyle{plain}

\section{Introduction}
There are three major sets of experimental results
that shed light on neutrino masses and mixings.
Solar-neutrino and reactor experiments have shown that
the electron neutrino \(\nu_e\) couples
to two nearly degenerate mass eigenfields
with \(10^{-10} \lsim 
m_1^2 - m_2^2 \lsim 10^{-3} \, \mathrm{eV}^2 \).
The atmospheric-neutrino experiments have shown
that the muon neutrino \(\nu_\mu\) and antineutrino \(\bar\nu_\mu\)
couple to two nearly degenerate mass eigenfields with 
\( 10^{-3} \lsim m_3^2 - m_4^2 
\lsim 10^{-2} \, \mathrm{eV}^2\)\@.
Double-beta-decay experiments have placed very stringent
limits on the rates of neutrinoless double beta decay.
\par
The current wisdom on neutrinos is that they have 
very small masses and that the tiny mass differences
experimentally observed are due to their small masses.
It is also generally believed that the smallness of the
masses of the neutrinos arises from the seesaw mechanism~\cite{mgmrm},
which involves huge Majorana masses.
It is also thought that CKM-like mixings explain
the solar and atmospheric data.
\par
The purpose of this paper is to present a general discussion
of neutrino masses and mixings and a rather different explanation
of the experimental facts.
It is argued that the neutrino mass matrix
is a \(6\times6\) complex symmetric matrix,
that the moduli of the Majorana mass terms
are small compared to those of the Dirac mass terms,
that neutrinos propagate as six mass eigenfields
with masses that form three nearly degenerate pairs,
and that the oscillations observed in the solar 
and atmospheric experiments are mainly between members of these pairs.
Neutrino masses and mixings have been discussed 
by several authors~\cite{Wolfenstein,PDauthors},
and the model introduced by Geiser~\cite{Geiser}
is an important example of the class of models
to be developed in what follows.
\par
Because left-handed fields can mix only with
other left-handed fields, neutrino masses
are best analyzed in terms of left-handed fields.
There are six such fields, the three left-handed flavor
eigenfields \(\nu_e, \nu_\mu\) and \(\nu_\tau\), and the
charge conjugates of the three right-handed neutrinos,
which are sterile.
In section 2 the neutrino mass matrix
is exhibited as a \(6\times6\) complex symmetric matrix \(\cM\)\@.
In general this matrix is not normal, but like every matrix it admits
a singular-value decomposition \(\cM = U M V^\dgr\)
in which the matrices \(U\) and \(V\) are unitary 
and the matrix \(M\) is real, nonnegative, and diagonal~\cite{Horn}.
Moreover because the matrix \(\cM\) is symmetric
it follows from a theorem of Takagi's~\cite{HornJohnson,Takagi}
that one may always choose \(V=U^*\)\@.
In section 3, Takagi's factorization \(\cM = U M U^\top\)
is used to diagonalize the action and 
to show that the singular values of the neutrino mass matrix \(\cM\)
or equivalently the diagonal elements of the matrix \(M\)
are the six neutrino masses \(m_j\) and that the unitary matrix \(U\)
describes the neutrino mixings.
\par
In section 4 the LEP measurement of the number of light
neutrino flavors is related to the mixing matrix \(U\)\@.
Section 5 describes a cosmological constraint on the masses 
of light, stable neutrinos. 
Section 6 expresses the neutrino oscillation probabilities
in terms of the mixing matrix \(U\) and the masses \(m_j\)
and briefly discusses the results of the solar, reactor,
and atmospheric neutrino experiments\@.
\par
In section 7 an angle \( x_\nu\) is introduced 
that quantifies the extent to which neutrinos are Dirac fermions
or Majorana fermions.  
Dirac neutrinos have \( x_\nu = 0\), and 
Majorana neutrinos have \( x_\nu = \pi/2\)\@.
If all Majorana mass terms vanish,
that is if \( x_\nu = 0\),
then the standard model conserves baryon-minus-lepton number,
\(B-L\), which is a global \(U(1)\) symmetry.
It is therefore natural in the sense of 't Hooft~\cite{tHooft}
to assume that \( x_\nu \approx 0 \) 
so that this symmetry is only slightly broken.
The neutrinos then are nearly Dirac fermions
and their masses coalesce into three pairs
of almost degenerate masses.
Thus the approximate conservation of \(B-L\)
explains the tiny mass differences
seen in the solar and atmospheric neutrino experiments
without requiring the
neutrino masses to be absurdly small.
There are three tiny mass differences,
of which one explains the solar experiments
and another the atmospheric ones, 
and three unconstrained mass differences,
which can lie in the eV range.
If one sets \( \sin x_\nu = 0.003\),
suppresses inter-generational mixing,
and imposes a quark-like mass hierarchy,
then one may fit the essential features of the
solar, reactor, and atmospheric neutrino data
with otherwise random mass matrices \(\cM\) in the eV range.
Thus neutrinos easily can have masses
that saturate the cosmological bound of about 8 eV\@.
Moreover because neutrinos are almost Dirac fermions,
neutrinoless double beta decay is suppressed by an extra factor
\(\sim \sin^2 x_\nu \, \sin^2 y_\nu \lsim 10^{-5}\),
where \( y_\nu\) is a second neutrino angle,
and is very slow,
with lifetimes in excess of \(2\times10^{27}\) years.
\par
An appendix outlines an efficient way of performing
the singular-value decomposition of the mass matrix \(\cM\)
by means of a call to the LAPACK~\cite{LAPACK} 
subroutine ZGESVD~\cite{lapack}\@.
\par
This \(B-L\) model of neutrino masses and mixings 
leads to these predictions about future experiments:
\begin{itemize}
\item The three flavor neutrinos oscillate mainly 
into the conjugates of the right-handed fields,
which are sterile, but the mass differences
among the three pairs of nearly degenerate masses
can lie in the eV range.  The probabilities of
the appearance of neutrinos or antineutrinos 
are small, as shown by LSND and KARMEN2,
but MiniBooNE has a reasonable chance of
seeing the appearance of \(\bar\nu_\mu\to\bar\nu_e\).
\item Because neutrino masses are not required
to be nearly as small as the solar and atmospheric
mass differences might suggest,
they may well be of cosmological significance.
\item If a suitable experiment can be designed,
it should be possible to see the tau neutrino disappear.
\item The rate of neutrinoless double beta decay           
is suppressed by an extra factor of 
\(\sim \sin^2 x_\nu \, \sin^2 y_\nu \lsim 10^{-5}\)
and hence will not be seen in the
Heidelberg/Moscow, IGEX, GENIUS, or CUORE
experiments. 
\end{itemize}

\section{Mass Terms and Mass Matrices}
There are three potential sources of neutrino masses.  
One source is Dirac mass terms,
which require the existence of right-handed neutrino fields.
The second source of neutrino masses
is Majorana mass terms
composed of left-handed neutrino fields;
these mass terms require a triplet of Higgs bosons,
break \(B-L\), and drive neutrinoless double beta decay. 
The third source is Majorana mass terms 
composed of right-handed neutrino fields;
these terms also break \(B-L\) but do not affect
neutrinoless double beta decay, at least in leading order.
Inasmuch as quarks and charged leptons come in both
left-handed and right-handed fields, the existence
of right-handed neutrino fields is reasonable.
And because the Higgs sector is still unknown,
the existence of Higgs bosons that transform as a triplet
under \(SU(2)_L\) remains possible.
So in this paper we shall consider all three kinds of mass terms.
\par
Because left- and right-handed fields transform differently
under Lorentz boosts, they cannot mix.
It is therefore convenient to write the
action density exclusively in terms of two-component, left-handed 
fields.  The two-component, left-handed neutrino
flavor eigenfields \( \nu_e, \nu_\mu, \nu_\tau\)
will be denoted \( \nu_i,\) for \(i = e, \mu, \tau\)\@.
The two-component, left-handed fields that are the
charge conjugates of the putative right-handed neutrino fields
\( n_{re}, n_{r\mu}, n_{r\tau}\) will be denoted
\( n_i = - i \s_2 \, n_{ri}^*\) for \(i = e, \mu, \tau \)
in which hermitian conjugation 
without transposition is denoted by an asterisk, and
\( \s_2 \) is the second Pauli spin matrix\@.
\par
For two left-handed fields \( \psi \) and \( \chi \),
there is only one mass term, and its hermitian form is
\beq
i \, m \chi^\top \s_2 \psi - i \, m^* \psi^\dgr \s_2 \chi^*
= i 
\sum_{\a=1}^2 \sum_{\b=1}^2 
\left( m \, \chi_\a \s_{2\a\b} \psi_\b - 
m^* \, \psi_\a^* \s_{2\a\b} \chi_\b^* \right),
\label {the mass term}
\eeq
which incidentally is invariant under a Lorentz
transformation \( g \in SL(2,C) \) because
the antisymmetric matrix \(\s_2\) converts
two factors of \(g\) into \( \det(g) = 1 \)\@.
If the field \( \chi \) is the charge conjugate
of a right-handed field 
that carries the same conserved quantum numbers
as the field \( \psi \), 
then this mass term (\ref{the mass term}) is called
a Dirac mass term.  
On the other hand, if the fields \( \psi \) and \( \chi \)
carry the same quantum numbers or no conserved
quantum numbers at all, then the mass term (\ref{the mass term})
is called a Majorana mass term.
Since electric charge is conserved,
only neutral fields can have Majorana mass terms.
\par
The six left-handed neutrino fields 
\( \nu_i, n_i \) for \( i=1, 2, 3\)
can have three kinds of mass terms.
The fields \( \nu_i \) and \( n_j \)
can form Dirac mass terms 
\beq
i D_{ij} \nu_i^\top \s_2 n_j - i D_{ij}^* n_j^\dgr \s_2 \nu_i^*
\label {Dirac mass term}
\eeq
in which the \(D_{ij}\) are complex numbers.
In a minimal extension of the standard model,
the \( D_{ij} \) are proportional to the mean value
in the vacuum of the neutral component of
the Higgs field.
The fields \( n_i \) and \( n_j \) can form
Majorana mass terms 
\beq
i E_{ij} n_i^\top \s_2 n_j - i E_{ij}^* n_j^\dgr \s_2 n_i^*.
\label {nn Majorana mass term}
\eeq
Within the standard model,
the complex numbers \( E_{ij} \)
are simply numbers;
in a more unified theory,
they might be the mean values
in the vacuum
of neutral components of Higgs bosons.
The fields \( \nu_i \) and \( \nu_j \)
can also form Majorana mass terms 
\beq
i F_{ij} \nu_i^\top \s_2 \nu_j - i F_{ij}^* \nu_j^\dgr \s_2 \nu_i^*.
\label {nunu Majorana mass term}
\eeq 
In a minimal extension of the standard model,
the complex numbers \( F_{ij} \) might
be proportional to the mean values
in the vacuum of the neutral component
of a new Higgs triplet \( h_{ab} = h_{ba} \)
that transforms as
\beq
h'_{ab}(x) = h_{a'b'} g_{a'a}^{-1}(x) g_{b'b}^{-1}(x)
\label {h'}
\eeq
in which the \( 2\times2 \) matrix \( g_{aa'}(x) \) is the one
that transforms the three doublets 
\beq
L_i(x) = \pmatrix{ \nu_i(x) \cr e_i(x) \cr }
\qquad \mathrm{as} \qquad
L_i(x)' = g(x) L_i(x)
\label {L}
\eeq
where \(e_i = e, \mu, \tau\) for \(i=1, 2, 3\)\@. 
Such a triplet Higgs multiplet would allow a term like
\beq
\sum_{a,b = 1}^2 h_{ab}(x)  L_{ia}(x)^\top \, \s_2 \, L_{jb}(x) 
\label {hterm}
\eeq
to remain invariant under \(SU_L(2)\otimes U(1)_Y\)
gauge transformations and so to contribute a mass term
to the action density.
\par
Since \( \s_2 \) is antisymmetric
and since any two fermion fields 
\( \chi \) and \( \psi \) anticommute,
it follows that 
\beq
\chi^\top \s_2 \psi = \psi^\top \s_2 \chi
\quad \mathrm{and}
\quad \chi^\dgr \s_2 \psi^* = \psi^\dgr  \s_2 \chi^* ,
\label {asb=bsa}
\eeq
which incidentally shows that a singlet Higgs field, 
\( h_{ab} = - h_{ba} \), 
would lead to \(F_{ij} = 0 \)\@.
This symmetry relation (\ref{asb=bsa}) implies that
the \(3 \times 3\) complex matrices \( E \) and \( F \)
are symmetric
\beq
E^{\top} = E \quad \mathrm{and} \quad F^{\top} = F
\label {symm complex}
\eeq
and that 
\beq 
i D_{ij} n_j^\top \s_2 \nu_i = i D_{ij} \nu_i^\top \s_2  n_j.
\label {DT}
\eeq
Let us introduce
the \( 6 \times 6 \) complex matrix \( \cM \) 
\beq
\cM = \pmatrix{ F & D \cr
                D^\top & E\cr}
\label {cM}
\eeq
which by (\ref{symm complex}) is symmetric.
Then with the six-vector \( N \) of left-handed neutrino fields
\beq
N = \pmatrix{\nu_e \cr \nu_\mu \cr \nu_\tau \cr 
n_e \cr n_\mu \cr n_\tau \cr}, 
\label {N}
\eeq
we may gather the mass terms 
into the matrix expression
\beq
\frac{i}{2} N^\top \cM \s_2 N 
- \frac{i}{2} N^\dgr \cM^* \s_2 N^*.
\label {the 6x6 mass term}
\eeq
Actually since each of the six fields in \(N\)
has two components, the six-vector \( N \) 
has twelve components.  The use of twelve-component
fields has been advocated by Rosen~\cite{SPQR} and more recently
by Starkman and Stojkovic~\cite{SS}\@.
\par
The complex symmetric
mass matrix \( \cM \) is not normal unless 
the positive hermitian matrix \( \cM \cM^\dgr \) is real
because the commutator \( [ \cM , \cM^\dgr ] \)
is twice its imaginary part:
\beq
[ \cM , \cM^\dgr ] = 2 i \, \Im m \left( \cM \cM^\dgr \right).
\label {abnormal}
\eeq
When the mass matrix is itself real, 
then it is also normal and may be diagonalized
by an orthogonal transformation.
But in general \( \cM \) is neither real nor normal.
\par
Yet like every matrix, 
the mass matrix \( \cM \)
possesses a singular-value decomposition~\cite{Horn,LAPACK}
\beq
\cM = U M V^\dgr
\label {SVD}
\eeq
in which the \(6\times6\) matrices \( U \) 
and \( V \) are unitary and the
\(6\times6\) matrix \( M \) is 
real and diagonal 
\beq
M = \pmatrix{m_1&0&0&0&0&0\cr
             0&m_2&0&0&0&0\cr
	     0&0&m_3&0&0&0\cr
	     0&0&0&m_4&0&0\cr
	     0&0&0&0&m_5&0\cr
	     0&0&0&0&0&m_6\cr}
\label {M}	     
\eeq
with singular values \( m_j \ge 0 \).
These singular values will turn out
to be the masses of the six neutrinos.
The columns of the unitary matrix \(U\)
are the left singular vectors of \(\cM\);
they are the eigenvectors of the matrix \(\cM\,\cM^\dgr\)
with eigenvalues \(m_j^2\)
\beq
\cM\,\cM^\dgr \, U = U \, M^2.
\label {Ueig}
\eeq
The columns of the unitary matrix \(V\)
are the right singular vectors of \(\cM\);
they are the eigenvectors of the matrix \(\cM^\dgr\,\cM\)
with eigenvalues \(m_j^2\)
\beq
\cM^\dgr\,\cM \, V = V \, M^2.
\label {Veig}
\eeq

\par
We have not yet made use of the fact that the complex
matrix \(\cM\) is symmetric.
It is a general mathematical theorem~\cite{HornJohnson}
due to Takagi~\cite{Takagi}
that for every symmetric complex matrix \(Z\)
there is a unitary matrix \(X\) and a real diagonal matrix \(D\)
with nonnegative elements such that \(Z = X D X^\top\)\@.
The diagonal matrix elements \(D_{jj}\) are the singular
values of the matrix \(Z\)\@.
Thus the symmetric complex matrix \(\cM\) may be factored 
by a unitary matrix \(W\) 
\beq
\cM = W M W^\top
\label {thm}
\eeq
where \(M\) is the matrix of masses (\ref{M})\@.
The columns of \(W\) are the eigenvectors of
the matrix \(\cM\,\cM^\dgr\) with eigenvalues \(m_j^2\)
\beq
\cM\,\cM^\dgr \, W = W \, M^2.
\label {Weig}
\eeq
Thus by (\ref{Ueig})
the columns of \(W\) are the same as the columns
of the unitary matrix \(U\) of the singular-value
decomposition (\ref{SVD})
apart from over-all phase factors.
Similarly the columns of \(W^*\)
are the eigenvectors of the matrix \(\cM^\dgr \,\cM\)
with eigenvalues \(m_j^2\)
\beq
\cM^\dgr \,\cM \, W^* = W^* \, M^2,
\label {W*eig}
\eeq
and so by (\ref{Veig})
the columns of \(W^*\) 
are the same as the columns
of the unitary matrix \(V\) of the singular-value
decomposition (\ref{SVD})
apart from over-all phases.
For our purposes the import of
Takagi's theorem is that
the singular-value decomposition
of the symmetric complex matrix \(\cM = U M V^\dgr\)
may be achieved with \(U=W\) and \(V=W^*\) so that
\beq
\cM = U \, M \, U^\top,
\label {Tak'sTh}
\eeq
which we may call Takagi's factorization.

\section{Field Equations, Masses, and Mixings}
The free, kinetic action density of 
a two-component left-handed spinor \( \psi \)
is \( i \psi^\dgr 
\left( \partial_0 - \vec \s \cdot \nabla \right)
\psi \)\@.
Thus by including the mass terms (\ref {the 6x6 mass term}),
one may write the free action density of
the six left-handed neutrino fields \( N \) as
\beq
\cL_0 = i N^\dgr 
\left( \partial_0 - \vec \s \cdot \nabla \right)
N
+ \frac{i}{2} N^\top \cM \s_2 N 
- \frac{i}{2} N^\dgr \cM^* \s_2 N^*
\label {N0ad1}
\eeq
or with Takagi's factorization (\ref{Tak'sTh})
of the mass matrix as
\beq
\cL_0 = i N^\dgr 
\left( \partial_0 - \vec \s \cdot \nabla \right)
N
+ \frac{i}{2} N^\top U \, M \s_2 \, U^\top \,N 
- \frac{i}{2} N^\dgr U^* \, M \s_2 U^\dgr \, N^*
\label {N0ad2}
\eeq
in which \(M\) is the real \(6\times6\)
diagonal matrix (\ref{M}) with nonnegative
entries \(m_j\)\@.
Let us define the six-vector of fields
\beq
N_m = U^\top N \quad \mathrm{or} \quad \nu_{m_i} = \sum_{j=1}^6 U_{ji} N_j.
\label {N_m=}
\eeq
Since the \(6\times6\) matrix \(U\) is unitary,
we have 
\beq 
N = U^*\,N_m \qquad \mathrm{and} \qquad
N^\top = N_m^\top \, U^\dgr
\label {NNt}
\eeq
as well as
\beq 
N^* = U\,N_m^* \qquad \mathrm{and} \qquad
N^\dgr = N_m^\dgr \, U^\top.
\label {NN*}
\eeq
Thus we may write \(\cL_0\) in the form
\bea
\cL_0 & = & i N_m^\dgr \, U^\top
\left(\partial_0 - \vec \s \cdot \nabla \right)
\, U^*  \, N_m \nn\\
& & \mbox{} + 
\frac{i}{2} N_m^\top \, U^\dgr \, U M \s_2 U^\top \, U^* \, N_m 
- \frac{i}{2} N_m^\dgr \, U^\top \, U^*  M \s_2 U^\dgr \, U \, N_m^*
\label {N0ad3}
\eea
or since \(U^\top \, U^* = I = U^\dgr \, U\) 
\beq
\cL_0 = i N_m^\dgr \,
\left(\partial_0 - \vec \s \cdot \nabla \right)
\, N_m 
+ \frac{i}{2} N_m^\top \, M \s_2 \, N_m 
- \frac{i}{2} N_m^\dgr \, M \s_2 \, N_m^*.
\label {N0ad4}
\eeq
The action density therefore has the diagonal form
\beq
\cL_0 = \sum_{i=1}^6 \left[ i \, \nu_{m_i}^\dgr \,
\left(\partial_0 - \vec \s \cdot \nabla \right)
\, \nu_{m_i} 
+ \frac{i}{2} m_i \left( \nu_{m_i}^\top \, \s_2 \, \nu_{m_i} 
- \nu_{m_i}^\dgr \, \s_2 \, \nu_{m_i}^* \right)\right].
\label {ADdiag}
\eeq
\par
The fields \(\nu_{m_i}\) are the normal modes of the theory.
They propagate according to the equation of motion
\beq
\left( \partial_0 - \vec \s \cdot \nabla \right)\, 
\nu_{m_i} 
= m_i \, \s_2 \, \nu_{m_i}^*
\label {dN=}
\eeq
in which the asterisk represents complex or hermitian
conjugation without transposition.
By taking the hermitian adjoint of this equation
without transposing the matrices and vectors
(or equivalently
by taking the hermitian adjoint of the equation
and then re-transposing the matrices and vectors),
one has
\beq
\left( \partial_0 - \vec \s^* \cdot \nabla \right)\, 
\nu_{m_i}^* 
= m_i \, \s_2^* \, \nu_{m_i}.
\label {dN*}
\eeq
The rules \( \s_2 \vec \s^* \s_2 = - \vec \s \),
\( (\s_2)^2 = 1 \), and \( \s^{2*} = - \s_2 \)
now imply that the equation of motion for \( \nu_{m_i}^* \) is
\beq
\left( \partial_0 + \vec \s \cdot \nabla \right)
\s_2 \nu_{m_i}^* = - m_i \, \nu_{m_i}.
\label {dN*=}
\eeq
Applying \( \left( \partial_0 + \vec \s \cdot \nabla \right) \)
to the field equation (\ref {dN=}) for \(\nu_{m_i}\)
and then using the field equation (\ref {dN*=}) for \(\nu_{m_i}^*\),
we find that
\beq
\left( \Box - m_i^2 \right) \nu_{m_i} = 0.
\label {TFE}
\eeq
Thus the diagonal elements \( m_i \) 
of the real mass matrix \(M\),
which are the singular values of the
mass matrix \( \cM \), are the neutrino masses,
and the eigenfield of mass \( m_i \) is
\(\nu_{m_i}\) as given by (\ref{N_m=})\@.
A neutrino of mass \( m_i \) described 
by the two-component field \(\nu_{m_i}\) 
must have two helicities or spin states
and so must be its own antiparticle.
The two-component field \(\nu_{m_i}\) is not hermitian,
however; we'd need to use four components
to make this field hermitian. 
\par
Since the mass eigenfields \(\nu_{m_i}\)
are the normal modes of the theory,
it is worth recoding their expansions
\beq
\nu_{m_i}(x) = \int\! \frac{d^3p}{(2\pi)^{3/2}}
\left[\,
u(\vec p, s, m_i) e^{ipx} a(\vec p,s,m_i) 
+ v(\vec p, s, m_i) e^{-ipx} a^\dagger (\vec p,s,m_i) \,
\right] 
\label {expansion}
\eeq
in which \( px = \vec p \cdot \vec x - e t \)\@.
In terms of the normalization factor 
\beq n_i = \frac{1}{2\sqrt{e(e+m_i)}},
\label {n}
\eeq
the two-component spinors \(u\) and \(v\) are
\beq
u(\vec p, {\thalf},m_i) = n_i \pmatrix{ m_i + e - p_3 \cr -p_1 -i p_2 \cr}
\quad
u(\vec p, -{\thalf},m_i) = n_i \pmatrix{ -p_1 +i p_2 \cr m_i + e + p_3 \cr} 
\label {us}
\eeq
and
\beq
v(\vec p, {\thalf},m_i) = n_i \pmatrix{ - p_1 +i p_2 \cr m_i + e + p_3 \cr}
\quad
v(\vec p, -{\thalf} ,m_i) = n_i \pmatrix{ p_3 - m_i -e \cr p_1 + ip_2 \cr}. 
\label {vs}
\eeq
They satisfy the spin sums
\beq
\sum_s u(\vec p, s, m_i) \, u^\dagger (\vec p, s, m_i)  = 
\sum_s v(\vec p, s, m_i) \, v^\dagger (\vec p, s, m_i)  = 
\frac{1}{2e} ( e - \vec p \cdot \vec \sigma )
\label {sumuuvv}
\eeq 
and 
\beq
\sum_s u(\vec p, s, m_i) v^\top (\vec p, s, m_i)  = 
\frac{i m_i}{2 e } \, \sigma_2.
\label {sumuv}
\eeq 
\par 
For momentum \( \vec p = p \hat z \) in the z-direction
with \( p \gg m_i \), these spinors reduce to 
\beq
u(\vec p, {\thalf},m_i) \approx \frac{m_i}{2p} \pmatrix{ 1 \cr 0 \cr}
\qquad
u(\vec p, -{\thalf},m_i) \approx \pmatrix{ 0 \cr 1 \cr} 
\label {pzus}
\eeq
and
\beq  
v(\vec p, {\thalf},m_i) \approx \pmatrix{ 0 \cr 1 \cr} 
\qquad
v(\vec p, -{\thalf}, m_i)\approx - \frac{m_i}{2p} \pmatrix{ 1 \cr 0 \cr},   
\label {pzvs}
\eeq
which shows that the field \(\nu_{m_i}(x)\) primarily
deletes neutrinos of negative helicity and adds
neutrinos of positive helicity. 
\par
In terms of the flavor eigenfields \(\nu_i\) and \(n_i\),
the eigenfield of mass \( m_i \) is
\beq
\nu_{m_j} = \sum_{i=1}^6 U_{ij} N_i = \sum_{i=1}^3 U_{ij} \nu_i
+ \sum_{i=1}^3 U_{(i+3)j} n_i.
\label {numj}
\eeq
The flavor eigenfields \( N \) are given by
\beq
N = U^* N_m.
\label {FEFs}
\eeq
In particular, the three left-handed fields
\( \nu_i \) for \( i = e, \mu, \tau \) are
linear combinations of the six mass eigenfields \( \nu_{m_j} \)
\beq
\nu_i = \sum_{j=1}^6 U_{ij}^* \nu_{m_j}
\label {nu=}
\eeq
and not simply linear combinations of three
mass eigenfields. 
The three active (right-handed) antineutrino
fields \( \nu_i^* \) for \( i = e, \mu, \tau \) 
are similarly
\beq
i \sigma_2 \nu_i^* = \sum_{j=1}^6 U_{ij} i \sigma_2 \nu_{m_j}^*. 
\label {nubar=}
\eeq
The matrix that expresses
the flavor eigenfields in terms of the mass eigenfields
is a \(3\times6\) matrix that is half of a \(6\times6\)
unitary matrix, not a \(3\times3\) unitary matrix.
The three sterile left-handed fields 
\( n_i \) for \( i = e, \mu, \tau \) are also 
linear combinations of the six mass eigenfields \( \nu_{m_j} \)
\beq
n_i = \sum_{j=1}^6 U_{(i+3)j}^* \nu_{m_j}.
\label {n=}
\eeq     

\par
In view of the relation (\ref{nu=}) between
the flavor eigenfields and the mass eigenfields,
the action density for 
neutrino interactions in four-component form~\cite{SWII} 
\beq
\cL' = 
\frac{-ie}{4\sin\theta_W}
\sum_{i=1}^3
\left\{
\sqrt{2} 
\left[
\bar e_i \not \!\! W ( 1 + \gamma_5 ) \nu_i
+ \bar \nu_i \not \!\! W^* 
( 1 + \gamma_5 ) e_i \right]
- \frac{\bar \nu_i \not \!\! Z ( 1 + \gamma_5 ) \nu_i}
{\cos\theta_W}
\right\}
\label {SWnu}
\eeq
may be written as
\bea
\cL' &\!=\!&
\frac{-ie}{4\sin\theta_W}
\sum_{i=1}^3
\left\{
\!\sqrt{2} 
\left[
\bar e_i \not \!\! W ( 1 + \gamma_5 ) 
\sum_{j=1}^6 U^*_{ij} \nu_{m_j} 
+ \sum_{k=1}^6 U_{ik} \bar \nu_{m_k}
\not \!\! W^*
( 1 + \gamma_5 ) e_i \right] \right.
\nn\\
&&\quad\qquad\qquad\mbox{} \left. - \sum_{j=1}^6 \sum_{k=1}^6
U_{ik} U^*_{ij} 
\frac{\bar \nu_{m_k} \not \!\! Z ( 1 + \gamma_5 ) \nu_{m_j}}
{\cos\theta_W}
\right\}.
\label {SWnum}
\eea
In two-component form, this is
\bea
\cL' &\!=\!&
\frac{e}{2\sin\theta_W}
\sum_{i=1}^3
\left\{
\sqrt{2}
\left[
e_i^\dgr ( W^0 + \vec W \cdot \vec \s )
\sum_{j=1}^6 U^*_{ij} \nu_{m_j} \right. \right.
\nn\\ 
&&\quad\qquad\qquad\mbox{} \left.
+ \sum_{k=1}^6 U_{ik} \nu_{m_k}^\dgr
( W^{0*} + \vec W^* \cdot \vec \s )
e_i \right] 
\nn\\
&&\quad\qquad\qquad\mbox{} \left. - \sum_{j=1}^6 \sum_{k=1}^6
U_{ik} U^*_{ij}
\frac{\nu_{m_k}^\dgr ( Z^0 + \vec Z \cdot \vec \s ) \nu_{m_j}}
{\cos\theta_W}
\right\}
\label {SWnum2}
\eea           
in which all fields are left-handed two-component spinors.

\section{LEP}
The four LEP measurements of
the invisible partial width
of the \(Z\)
impose upon the number of
light neutrino types the constraint~\cite{LEP}
\beq
N_\nu = 2.984 \pm 0.008.
\label {Nnu}
\eeq
\par
It follows from the action density (\ref{SWnum})
that the amplitude for the \(Z\) production of two
neutrinos \( \nu_{m_j} \) and \( \nu_{m_k} \) to lowest order
is 
\beq
A(\nu_{m_j}, \nu_{m_k} ) \, \propto \sum_{i=1}^3 U_{ik} U^*_{ij},
\label {A}
\eeq
and therefore that the cross-section for that process is 
\beq
\s(\nu_{m_j}, \nu_{m_k} ) \, \propto | \sum_{i=1}^3 U_{ik} U^*_{ij} |^2.
\label {x}
\eeq
The measurement (\ref {Nnu}) 
of the number \( N_\nu \) of light neutrino species
thus implies that the sum over the light-mass eigenfields is
\beq
\sum_{j,k \; \mathrm{light}}
| \sum_{i=1}^3 U_{ik} U^*_{ij} |^2
= 2.984 \pm 0.008. 
\label {constraint}
\eeq
\par
This constraint on the \(6\times6\) unitary matrix \(V\)
is quite well satisfied if all six neutrino masses
are light.  For in this all-light scenario, the sum is
\beq
\sum_{j,k=1}^6 \sum_{i=1}^3 \sum_{i'=1}^3
U_{ik} U^*_{ij}
U^*_{i'k} U_{i'j}
= \sum_{i=1}^3 \sum_{i'=1}^3 
\d_{ii'} \d_{ii'}
= \sum_{i=1}^3 1 = 3 
\approx 2.984 \pm 0.008.
\label {all light}
\eeq
But the constraint (\ref {Nnu}) will also be satisfied
if the three flavor eigenfields \( \nu_i \)
couple only to light-mass eigenfields \( \nu_{m_j} \),
for in this case the matrix elements \( U^*_{ij} \) 
between flavor \(i\) and heavy mass \(m_j\) vanish,
and one may extend the sum over
the light-mass eigenfields to a sum over all six
mass eigenfields.
In this few-light scenario,
the three independent flavor eigenfields \( \nu_i \)
must couple to at least three light-mass eigenfields \( \nu_{m_j} \).

\section{Cosmology}
\subsection{Weighing Neutrinos} 
Any stable, two-component neutrino mass eigenfield \(\nu_{m_j}\) 
that couples via \( U^*_{ij}\) to a flavor eigenfield
must have a contemporary density of \(115/\mathrm{cc}\)
due to thermal equilibrium in the early universe~\cite{FKM,Kolb}.
Thus the present mass density of the 
eigenfield \(\nu_{m_j}\) is \(115 \, m_j /\mathrm{cc}\).
The current critical density is 
\(1.05 \times 10^4 \, h^2 \, \mathrm{eV/cc}\),
where \(h\) is the Hubble parameter in units
of 100 km/sec/Mpc.
Thus if \( h \approx 0.65\),
then the contribution of \( \nu_{m_j} \)
to the neutrino critical density \(\Omega_\nu\) 
is \( m_j / 40 \, \mathrm{eV} \).
So the contribution to \(\Omega_\nu\)
of the stable, light mass eigenfields \(\nu_{m_j}\)
that couple to the three flavor neutrinos is
\beq
\Omega_\nu = \left( \sum_{j \; \mathrm{light}} m_j \right) 
\frac{1}{40\,\mathrm{eV}}.
\label {Onu}
\eeq
Since \( \Omega_\nu \) is the neutrino contribution
to Hot Dark Matter, and since the HDM part of \(\Omega_M\)
is probably less than 0.2, we may conservatively conclude
that \( \Omega_\nu \lsim 0.2 \)\@.
Thus we arrive at an approximate upper
bound~\cite{FKM,Kolb} for the sum of the masses \( m_j \) 
of the light, stable neutrinos that couple to the three flavor
eigenfields:
\beq
\sum_{j \; \mathrm{light}} m_j \, \lsim \, 8 \, \mathrm{eV}.
\label {eubound}
\eeq
\par
Under other assumptions Fukugita, Liu, and Sugiyama~\cite{FLS}
have derived limits on the sum of the light neutrino masses
that are in the range of 2 to 5 eV\@.
The Sloan Digital Sky Survey~\cite{SDSS} will measure  
\(\Omega_\nu\) and weigh the light neutrinos~\cite{HET}\@.
Within the context of a minimally extended
standard model,
these cosmological bounds
and the LEP constraint (\ref{constraint}) 
imply that the \(Z\) gauge boson couples
to between 3 and 6 very light neutrinos.
\subsection{BBN Bounds on Light Sterile Neutrinos}
\par
It is open question whether active neutrinos
oscillate mainly into other active neutrinos
or mainly into sterile neutrinos. 
If they oscillate mainly into sterile neutrinos,
then in the early universe electron neutrinos 
before they decoupled might have brought sterile
neutrinos into chemical equilibrium~\cite{BBNlimit},
increasing the effective number \(N^\mathrm{eff}_\nu\) 
of two-component neutrino species.
A higher \(N^\mathrm{eff}_\nu\)  
would have raised the energy density \(\rho\) of the universe
and therefore its rate \(\dot R\) of expansion
since \( \dot R^2 = (8\pi G/3) \rho R^2 - k \)
where \(k\) is the curvature.
The neutron-to-proton ratio would then have 
been higher at its freeze-out (\(T \approx 0.7 \) MeV),
and the abundance \(Y\) of helium higher 
than is observed (\(Y \approx 0.24\pm0.01\)). 
In the case of maximal mixing between \(\nu_\mu\)
and \(\nu_{\mu s}\),
the conservative BBN limit \(N^\mathrm{eff}_\nu < 4\) 
would impose on 
\(\delta m^2_{\nu_\mu} = | m^2_{\nu_\mu} - m^2_{\nu_{\mu s}}|\)
an upper limit of the order of \(10^{-6}\) eV\(^2\)~\cite{BBNlimit}.
This big-bang-nucleosynthesis (BBN) constraint 
would rule out the possibility that
the atmospheric muon neutrinos oscillate mainly
into sterile muon neutrinos.

\subsection{Lepton Asymmetries} 
However an excess of \(\nu_e\) over \(\bar \nu_e\),
or \textit{vice versa},  
when the temperature was between 30 and 0.7 MeV,
would have modified the reaction rates
\(n+\nu_e \leftrightarrow p + e^-\)
and \(n+e^+ \leftrightarrow p + \bar \nu_e\) 
and consequently changed the effective number
of neutrino species \(N^\mathrm{eff}_\nu \)\@. 
Thus Foot and Volkas have pointed out~\cite{FootVolkas} 
that the BBN constraints on active-sterile
neutrino oscillations 
depend upon the assumption that the
lepton-number asymmetry of the early universe
was negligible.
They also have argued~\cite{FootVolkas} that active-sterile
neutrino oscillations could themselves have generated 
a significant lepton-number asymmetry 
at the relevant temperatures.
So the BBN limits on \(N^\mathrm{eff}_\nu \)
and \(\delta m^2_{\nu_\mu}\)
may not be valid.
And the recent measurement
of the cosmic microwave background (CMB) radiation
by the Boomerang experiment
has called into question the precision of BBN results;
it now appears that the BBN bound
on the baryon density ought to be raised by about 50\%
to \( 0.024 < h^2 \Omega_b < 0.042\)~\cite{MaxMatias}.
\par
The Affleck-Dine mechanism~\cite{AD} has been 
suggested as a second way of generating \(B\) and \(L\) 
asymmetries in the minimal supersymmetric model~\cite{MD}
and its extensions~\cite{CCG}. 

\subsection{Baryon Number of the Universe}
K.~Dick, M.~Lindner, M.~Ratz, and D.~Wright~\cite{Wright} 
recently showed how a universe that conserves \(B-L\)
might have started with \(B=L=0\) and still have arrived 
at the observed current baryon-to-photon ratio  
as long as the masses of the neutrinos are suitably small.
In their model a density of \(B+L\) produced
at \(GUT\) temperatures is equilibrated between 
left- and right-handed (then massless) particles by 
Yukawa processes at temperatures above 1 TeV\@. 
Sphalerons~\cite{sphalerons} then washed out 
left-handed baryons and leptons,
reducing \(B+L\) while conserving \(B-L\)\@.
If the Yukawa couplings of the neutrinos
are so weak as to yield (purely Dirac) neutrino masses 
less than about 10 keV, 
then the equilibration of left- and right-handed neutrinos 
could not have kept up with the sphaleron washout, 
and an excess of left-handed baryons 
equal to the excess of right-handed neutrinos,
which are immune to sphaleron washout, 
would have survived the electroweak era. 
This mechanism is both a potential explanation
of the baryon number of the universe
and a third way of generating 
lepton asymmetries.
\par
Since \(B-L\) is exactly conserved in this model,
neutrino oscillations occur only among the three
active flavors.  But this model might work
even if \(B-L\) were slightly broken. 

\section{Oscillations}
It follows from the action density (\ref {SWnum})
and from arguments presented elsewhere~\cite{Boris}
that the lowest-order amplitude \(A(\nu_i \to \nu_{i'})\)
for a neutrino \(\nu_i\)
(\textit{e.g.,} produced by a charged lepton \(e_i\))
after propagating with energy \(E\)
a distance \(L\) as some light-mass eigenfield
of mass \( m_j \ll E \) to appear as
a neutrino \(\nu_i'\)
(\textit{e.g.,} producing a charged lepton \(e_i'\)) is
\beq
A(\nu_i \to \nu_{i'}) =
\sum_{j\;\mathrm{light}}
U^*_{i'j} U_{ij}
e^{-im_j^2 L/(2E)}. 
\label {nunuosc}
\eeq
In view of (\ref{nubar=}),
the lowest-order amplitude for the 
anti-process, \( \bar \nu_i \to \bar \nu_{i'} \), 
is 
\beq
A(\bar \nu_i \to \bar \nu_{i'}) =
\sum_{j\;\mathrm{light}}
U_{i'j} U^*_{ij}
e^{-im_j^2 L/(2E)},
\label {barnunuosc}
\eeq
which is \(A(\nu_i \to \nu_{i'})\) with 
the matrix elements of \(U\)
replaced by their complex conjugates~\cite{Cabibbo}.                           \par    
The corresponding probabilities 
to lowest order are 
\beq
P(\nu_i \to \nu_{i'}) = 
\sum_{j,j'\,\mathrm{light}}
U^*_{i'j} U_{ij} U_{i'j'} U^*_{ij'} 
\exp{\left[i(m_{j'}^2 - m_{j}^2)L/(2E)\right]} 
\label {Pnunu'}
\eeq
and 
\beq
P(\bar\nu_i \to \bar\nu_{i'}) = 
\sum_{j,j'\,\mathrm{light}}
U^*_{i'j} U_{ij} U_{i'j'} U^*_{ij'} 
\exp{\left[i(m_j^2 - m_{j'}^2)L/(2E)\right]}
\label {Pnunu'bar}
\eeq
both of which, 
if all six neutrinos are light,
approach \(\d_{i i'}\) 
in the limit \(L/E \to 0\)\@.
\par
A measure of \(CP\) violation is provided by
the asymmetry parameter \(A(i,i')\) defined as 
\beq
A(i,i') = \frac{P(\nu_i \to \nu_{i'}) - P(\bar\nu_i \to \bar\nu_{i'})} 
{P(\nu_i \to \nu_{i'}) - P(\bar\nu_i \to \bar\nu_{i'})}
\label {Asym} 
\eeq
and given by
\beq
A(i,i') = \frac{\sum_{j,j'\,\mathrm{light}}
\mathit{Im} \left(U^*_{i'j} U_{ij} U_{i'j'} U^*_{ij'}\right) 
\sin{\left[(m_j^2 - m_{j'}^2)L/(2E)\right]} } 
{\sum_{j,j'\,\mathrm{light}}
\mathit{Re} \left(U^*_{i'j} U_{ij} U_{i'j'} U^*_{ij'}\right) 
\cos{\left[(m_j^2 - m_{j'}^2)L/(2E)\right]} } . 
\label {Asymis}
\eeq 
\par
As we shall see in the next section,
if the six neutrinos are light and either purely Dirac 
or purely Majorana,
then for each \(i = e, \mu, \tau\) the sum over \(i'\) is unity:
\beq
\sum_{i'=e}^\tau P(\nu_i \to \nu_{i'}) = 1;
\label {dirsum1}
\eeq
but if the neutrinos are nearly but not quite Dirac fermions,
then this sum of probabilities tends to be about
\beq
\sum_{i'=e}^\tau P(\nu_i \to \nu_{i'}) \approx \half.
\label {ndirsum.5}
\eeq 
\par
If for simplicity we stretch the error bars
on the Chlorine experiment,
then the solar neutrino experiments,
especially Gallex and SAGE,
show a diminution of electron neutrinos
by a factor of about one-half:
\beq
P_{\mathrm{sol}}(\nu_e \to \nu_e ) \approx \half,
\label {solexp}
\eeq 
which requires a pair of mass eigenstates
whose squared masses differ 
by at least \(\sim 10^{-10} \, \mathrm{eV}^2\)~\cite{FKM}\@.
The reactor experiments, Palo Verde
and especially Chooz, imply that
these squared masses differ by less than
\( \sim 10^{-3} \, \mathrm{eV}^2\)~\cite{FKM}\@.
\par
The atmospheric neutrino experiments,
Soudan II, Kamiokande III, IMB-3, and
especially SuperKamiokande,
show a diminution of muon neutrinos
and antineutrinos by about one-third:
\beq
P_{\mathrm{atm}}(\nu_\mu \to \nu_\mu ) \approx \frac{2}{3},
\label {atmexp}
\eeq 
which requires a pair of mass eigenstates 
whose squared masses differ by 
\(10^{-3} \lsim |m_j^2 - m_k^2| 
\lsim 10^{-2} \, \mathrm{eV}^2\)~\cite{FKM}\@.   
\par
If the LSND neutrino experiment is correct,
then it requires a pair of states
whose squared masses differ by 
\(10^{-1} \lsim |m_j^2 - m_k^2| \lsim
10^{+1} \, \mathrm{eV}^2\)~\cite{FKM}\@.

\section{The \(B-L\) Model}
When the Majorana mass matrices \(E\) and \(F\)
are both zero, the action density (\ref {N0ad1})
is invariant under the \(U(1)\) transformation
\beq
N' = e^{i \theta G} \, N
\label {U1}
\eeq
in which the \(6\times6\) matrix \(G\)
is the block-diagonal matrix
\beq
G = \pmatrix{ I & 0 \cr
              0 & -I \cr}
\label {G}
\eeq
with \( I \) the \(3\times3\) identity matrix.
The kinetic part of (\ref {N0ad1}) is clearly
invariant under this transformation.
The mass terms are invariant only when
the anti-commutator
\beq
\{ \cM, G \} = 2 \pmatrix{ F & 0 \cr 0 & - E \cr} = 0
\label {0}
\eeq
vanishes.
\par
This \(U(1)\) symmetry is 
the restriction to the neutrino sector
of the symmetry generated by 
baryon-minus-lepton number, \(B-L\), 
which is exactly conserved in the standard model.
A minimally extended standard model
with right-handed neutrino fields \( n_{ri} \)
and a Dirac mass matrix \(D\) but with
no Majorana mass matrices,
\(E = F = 0\), also conserves \(B-L\)\@.
When \(B-L\) is exactly conserved, 
\emph{i.e.,} when \( D \ne 0 \) but \(E = F = 0\),
then the six neutrino masses \(m_j\)
collapse into three pairs of degenerate masses
which may be combined into three 
Dirac neutrinos. 
\par
Suppose this symmetry is slightly broken by the Majorana
mass matrices \(E\) and \(F\).
Then for random mass matrices \( D \), \(E\), and \(F\),
the six neutrino masses \(m_j\) will form three
pairs of nearly degenerate masses
as long as the ratio
\beq
\sin^2 x_\nu = \frac{\mathrm{Tr}( E^\dgr E + F^\dgr F )}
{\mathrm{Tr}( 2 D^\dgr D + E^\dgr E + F^\dgr F )}
\label {s}
\eeq
is small.   
For a generic mass matrix \(\cM\),
the parameter \(\sin^2 x_\nu\) lies between the extremes
\beq
0 \le \sin^2 x_\nu \le 1
\label {0s1}
\eeq
and characterizes the kind of the neutrinos.
The parameter \(\sin^2 x_\nu\) is zero for purely Dirac neutrinos
and unity for purely Majorana neutrinos.
\par
Let us now recall 't Hooft's definition~\cite{tHooft}
of naturalness: 
It is natural to assume that a parameter is small
if the theory becomes more symmetrical when the parameter vanishes.
In this sense it is natural to assume that 
the parameter \( \sin^2  x_\nu \)
is small because the minimally extended standard model
becomes more symmetrical, conserving \(B-L\),
when \( \sin^2  x_\nu = 0 \)\@.
\par
\begin{figure}
\centering
\includegraphics{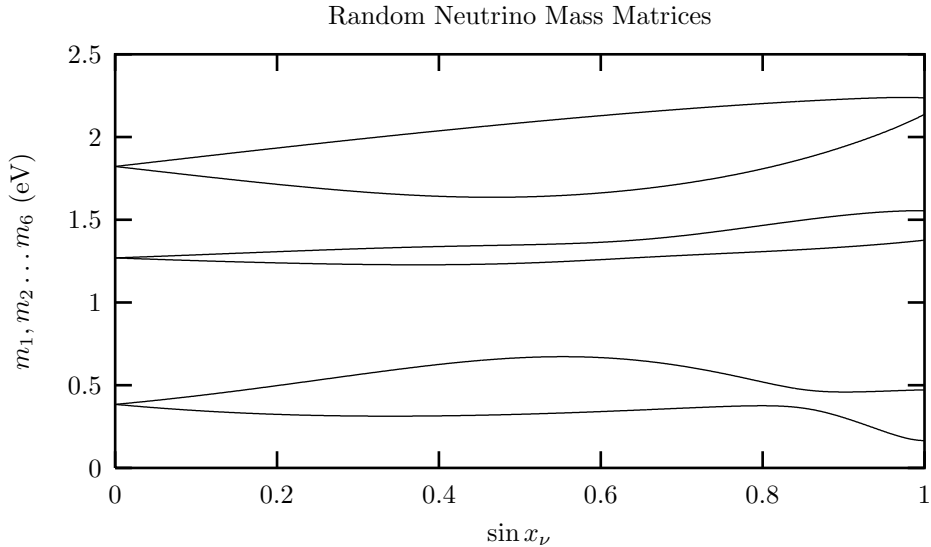}
\caption{The six neutrino masses are plotted against
the parameter \(\sin x_\nu\) for a set of 
random \(6\times6\) mass matrices.}
\label{fig:g1}
\end{figure}

In Fig.~\ref{fig:g1} the six neutrino masses \(m_j\)
are plotted for a set of mass matrices \(\cM\)
that differ only in the parameter \(\sin x_\nu\).
Apart from \(\sin x_\nu\),
every other parameter of the mass matrices \(\cM\) 
is a complex number \(z = x + i y \) in which
\(x\) and \(y\) were chosen
randomly and uniformly on the interval
\([ -1 \, \mathrm{eV}, 1 \, \mathrm{eV} ]\)\@.
It is clear in the figure that when \( \sin^2 x_\nu \approx 0 \),
the six neutrino masses \(m_j\) coalesce into three nearly degenerate pairs.
Although the six masses of the neutrinos are in the eV range,
they form three pairs with very tiny mass differences
when \( \sin^2 x_\nu \approx 0 \)\@.
\par
Thus the very small mass differences required by the
solar and atmospheric experiments are naturally
explained by the assumption that the symmetry
generated by \(B-L\)
is broken only slightly by the Majorana
mass matrices \(E\) and \(F\)\@. 
This same assumption implies that neutrinos are
very nearly Dirac fermions and hence explains
the very stringent upper limits on neutrinoless
double beta decay~\cite{FKM}.
Because the masses of the six neutrinos may lie
in the range of a few eV,
instead of being squashed down to
the meV range by the seesaw mechanism,
they may contribute to hot dark matter 
in a way that is 
cosmologically significant.
This \(B-L\) model with \( \sin^2 x_\nu \approx 0 \)
is the converse of the seesaw mechanism. 
\subsection*{An Example}
\par
A simple illustration of these ideas is provided
by the mass matrix \(\cM\) 
\beq
\cM = \pmatrix{f_1\sin x&0&0&d_1\cos x&0&0\cr
             0&f_2\sin x&0&0&d_2\cos x&0\cr
             0&0&f_3\sin x&0&0&d_3\cos x\cr
             d_1\cos x&0&0&e_4\sin x&0&0\cr
             0&d_2\cos x&0&0&e_5\sin x&0\cr
             0&0&d_3\cos x&0&0&e_6\sin x\cr},
	     \label {Matrix}
\eeq 
in which all the elements are taken to be real and non-negative.
At \( x=\pi/2\), the Majorana parameter \( \sin x_\nu \)
is unity, and the neutrinos are purely Majorana.
The six flavor eigenfields \( \nu_1, \nu_2, \nu_3,
n_1, n_2, n_3\) are then also mass eigenfields
with masses \( f_1, f_2, f_3, e_1, e_2, e_3\), respectively.
\par
For \( x\) between \(\pi/2\) and zero,
the fields \( \nu_i \) and \( n_i \) for each \(i=1,2,3\)
mix to form three pairs of mass eigenfields with masses
\beq
m_{i\pm} = \frac{1}{2}\,
\left| (e_i+f_i)\sin x \pm \sqrt{ (e_i-f_i)^2\sin^2 x
+ 4 d_i^2 \cos^2 x } \right|,
\label {exmass}
\eeq
which are the singular values of \(\cM\)\@.
The difference of the squared masses of the \(i\)th pair
is proportional to \(\sin x\)
\beq
m_{i+}^2 - m_{i-}^2
= (e_i+f_i) \sqrt{ (e_i-f_i)^2\sin^2 x
+ 4 d_i^2 \cos^2 x } \, \sin x.
\label {exdm2}
\eeq
If the mass parameters, \(f_i, e_i, d_i\), are comparable
in magnitude, then the Majorana parameter \( \sin x_\nu \)
is roughly \( \sin x\)\@. 
\par
At \( x=0\), \( x_\nu=0\), 
the neutrinos are purely Dirac fermions,
and the pair of mass eigenfields 
\beq
\nu_{m_i+} = \frac{1}{\sqrt{2}}\,(\nu_i +  n_i )
\quad \mathrm{and} \quad
\nu_{m_i-} = \frac{-i}{\sqrt{2}}\,(\nu_i -  n_i )
\label {dexmass}
\eeq
have the same mass or singular value,
\(m_i = |m_{i\pm}| = d_i\)\@.
Since these fields are degenerate,
the linear combinations
\beq
\nu_i = \frac{1}{\sqrt{2}}\,(\nu_{m_i+} + i \nu_{m_i-} )
\quad \mathrm{and} \quad
n_i = \frac{1}{\sqrt{2}}\,(\nu_{m_i+} - i \nu_{m_i-} )
\label {partsdiracnu}
\eeq
are also mass eigenfields;
they may be combined to form a single Dirac neutrino
\beq
\psi_i = \pmatrix{\nu_i\cr i \s_2 n_i^*\cr}
\label {diracnu}
\eeq
of mass \(m_i\) for each \(i=1,2,3\)\@.
It follows from Eqs.~(\ref{dN=}) and (\ref{dN*=})
that \(\psi_i\) satisfies Dirac's equation 
\((\gamma^n\partial_n + m_i)\psi_i=0\)\@.
When \(x_\nu\) is not zero or near zero,
then there is no reason why any two of the 
masses \(m_i\) should be equal, and one may
not combine two of the fields to form a Dirac neutrino.
\par
For very small values of the angle \( x\)
and of the parameter \( \sin^2 x_\nu \),
the neutrinos are nearly Dirac,
and the three squared-mass differences
\beq
m_{i+}^2 - m_{i-}^2 
\approx 2 d_i (e_i+f_i) \cos  x \sin  x
\label {dm2ap}
\eeq
being proportional
to \( \sin x \approx \sin x_\nu \),
are very small,
and the mass eigenfields are approximately
\beq
\nu_{m_i\pm} \approx \frac{1}{\sqrt{2}}\,(\nu_i \pm  n_i )
+ \frac{e_i - f_i}{2\sqrt{2} \, d_i} \, \tan  x \, n_i.
\label {ndapmass}
\eeq
\par
If \( \sin^2 x_\nu = 0 \), then there are 
three purely Dirac neutrinos, 
and the mixing matrix \(V\) is block diagonal
\beq
V = \pmatrix{ u^* & 0 \cr 0 & v \cr}
\label {VD}
\eeq
in which the \(3\times3\) unitary matrices \(u\) and \(v\)
occur in the singular-value decomposition of the
\(3\times3\) matrix \(D = u \, m \, v^\dgr \)\@.
If these three Dirac neutrinos are also light, 
then unitarity implies that 
the sum of the normalized probabilities is unity
\beq
\sum_{i'=e}^\tau P(\nu_i \to \nu_{i'}) = 1.
\label {dirsum1.}
\eeq
If \( \sin^2 x_\nu = 1 \), then
this sum is also unity by unitarity 
because in this case the mixing matrix for the six
purely Majorana neutrinos is
\beq
V = \pmatrix{ v_F & 0 \cr 0 & v_E \cr}.
\label {VM}
\eeq
But if there are six light, nearly Dirac neutrinos,
then each neutrino flavor \(\nu_i\) will 
oscillate both into other neutrino flavor eigenfields
and into sterile neutrino eigenfields.
In this case this sum tends to be about a half
\beq
\sum_{i'=e}^\tau P(\nu_i \to \nu_{i'}) \approx \half.
\label {gensum.5}
\eeq 
\subsection*{Fitting the Solar and Atmospheric Data}
\par
\begin{figure}
\centering
\includegraphics{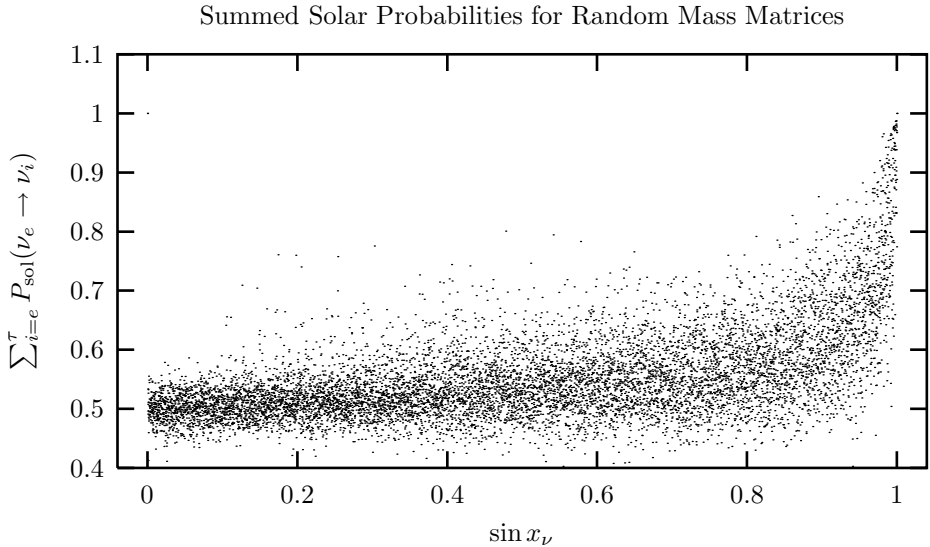}
\caption{The sum of the
probabilities \( P_{\mathrm{sol}}(\nu_e \to \nu_i ) \) 
for solar neutrinos summed over \(i\)
for \(i = e, \mu, \tau\)
for 10000 random mass matrices \(\cM\)
are plotted as a function
of the parameter \(\sin x_\nu\)\@.}
\label{fig:s2}
\end{figure}
\begin{figure}
\centering
\includegraphics{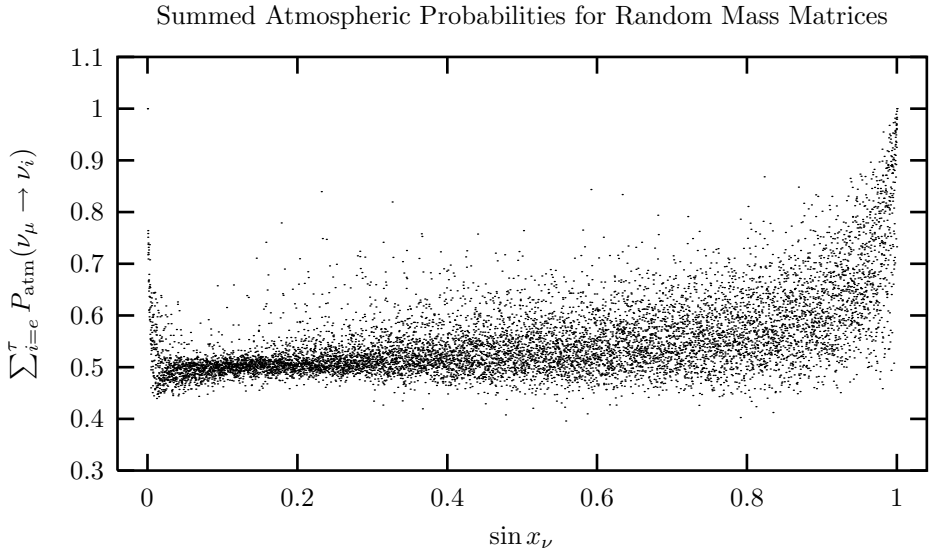}
\caption{The sum of the
probabilities \( P_{\mathrm{atm}}(\nu_\mu \to \nu_i ) \) 
for atmospheric neutrinos summed over \(i\)
for \(i = e, \mu, \tau\)
for 10000 random mass matrices \(\cM\)
are plotted as a function
of the parameter \(\sin x_\nu\)\@.}
\label{fig:h2}
\end{figure}

This intuition is supported by Figs.~\ref{fig:s2}
and \ref{fig:h2} which respectively
display for 10000 random mass matrices \(\cM\) the sums 
of probabilities
\( \sum_{i'=e}^\tau P_{\mathrm{sol}} (\nu_e \to \nu_{i'}) \)
and
\( \sum_{i'=e}^\tau P_{\mathrm{atm}} (\nu_\mu \to \nu_{i'}) \)
as a function of the parameter \(\sin x_\nu\)\@.
The plots clearly show that these sums tend to cluster
around the value \(\half\) when \( \sin x_\nu \)
is small but not infinitesimal.
The points at \( \sin x_\nu = 0 \) 
and at \( \sin x_\nu = 1 \) display
the unitarity relation (\ref {dirsum1.})
for respectively purely Dirac and purely Majorana neutrinos.
\par
In these scatter plots and in those that follow,
every parameter of the 10000 matrices is a complex number
\( z = x + i y \) with \(x\) and \(y\) chosen
randomly and uniformly from the interval \( [ - 1 \mathrm{eV},
1 \mathrm{eV} ] \)\@.
The solar neutrinos are taken to have an energy of 1 MeV,
and the probability (\ref {Pnunu'}) is averaged 
over one revolution of the Earth about the Sun.
The atmospheric neutrinos are averaged over the atmosphere
and over energies in the range of 1--30 GeV
weighted by the flux of atmospheric muon neutrinos
as a function of energy and local zenith angle
given by the Bartol group in Table V of ref.~\cite{Gaisser}\@.
\par
\begin{figure}
\centering
\includegraphics{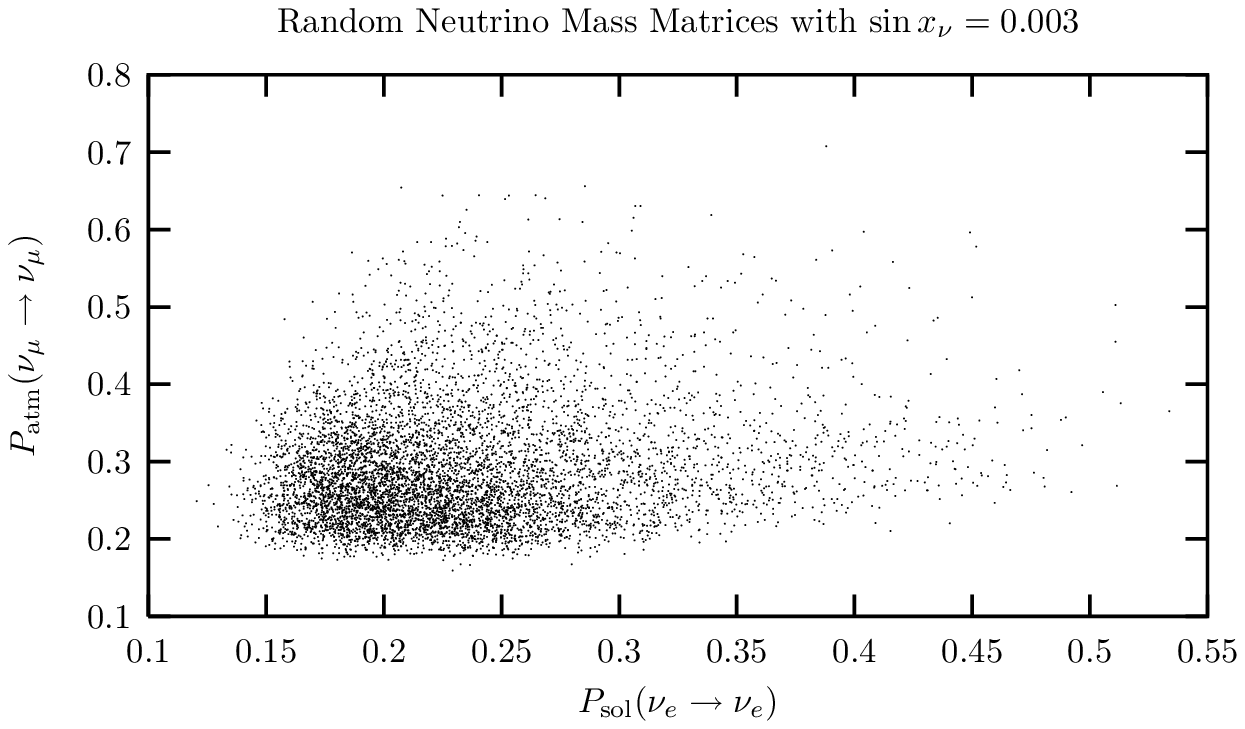}
\caption{The probabilities \( P_{\mathrm{sol}}(\nu_e \to \nu_e ) \)
and \( P_{\mathrm{atm}}(\nu_\mu \to \nu_\mu ) \) 
for 10000 random mass matrices \(\cM\)
all with the parameter \(\sin x_\nu = 0.003\)\@.}
\label{fig:e4}
\end{figure}

If we continue to interpret the Chlorine experiment
very conservatively,
then the data of the solar and atmospheric experiments 
(\ref{solexp}) and (\ref{atmexp}) require
that 
\beq
P_{\mathrm{sol}}(\nu_e \to \nu_e ) \approx \half 
\qquad \mathrm{and}\qquad
P_{\mathrm{atm}}(\nu_\mu \to \nu_\mu ) \approx \frac{2}{3}.
\label {bothexp}
\eeq
But because of the approximate, empirical sum rule (\ref {gensum.5})
for \(i = e\) and \(\mu\),
the only way in which the probabilities
\( P_{\mathrm{sol}}(\nu_e \to \nu_e ) \)     
and \( P_{\mathrm{atm}}(\nu_\mu \to \nu_\mu ) \)
can fit these experimental results
is if inter-generational mixing is suppressed
so that \( \nu_e \) oscillates into
\(n_e\) and so that \(\nu_\mu\)
oscillates into \(n_\mu\)\@.
In other words,
random mass matrices \( \cM \),
even with \( \sin x_\nu \approx 0 \),
produce probabilities \( P_{\mathrm{sol}}(\nu_e \to \nu_e ) \)
and \( P_{\mathrm{atm}}(\nu_\mu \to \nu_\mu ) \)
(suitably averaged respectively over the Earth's orbit 
and over the atmosphere) that are too small.
The reason is that for small \( \sin x_\nu \)
the flavor eigenfield \(\nu_e\) or \(\nu_\mu\)
necessarily is split into two mass eigenfields,
which reduces the probabilities to \(\sim \half\);
and so, if there is further inter-generational mixing,
then these probabilities tend to be too small.
This effect is illustrated in Fig.~\ref{fig:e4}
in which the probabilities
\( P_{\mathrm{sol}}(\nu_e \to \nu_e ) \)
and \( P_{\mathrm{atm}}(\nu_\mu \to \nu_\mu ) \) 
are plotted for 
10000 \footnote{Due to the arXiv limit
of 650 kb on the length of articles, in some of the figures
only 6000 points are plotted.} 
random mass matrices
\(\cM\) all with \( \sin x_\nu = 0.003 \)\@.
Figure~\ref{fig:e4}
makes it clear that 
the probabilities \( P_{\mathrm{sol}}(\nu_e \to \nu_e ) \)
and \( P_{\mathrm{atm}}(\nu_\mu \to \nu_\mu ) \) are too small
to fit the experimental results (\ref{solexp}) and (\ref{atmexp}).  
\par
\begin{figure}
\centering
\includegraphics{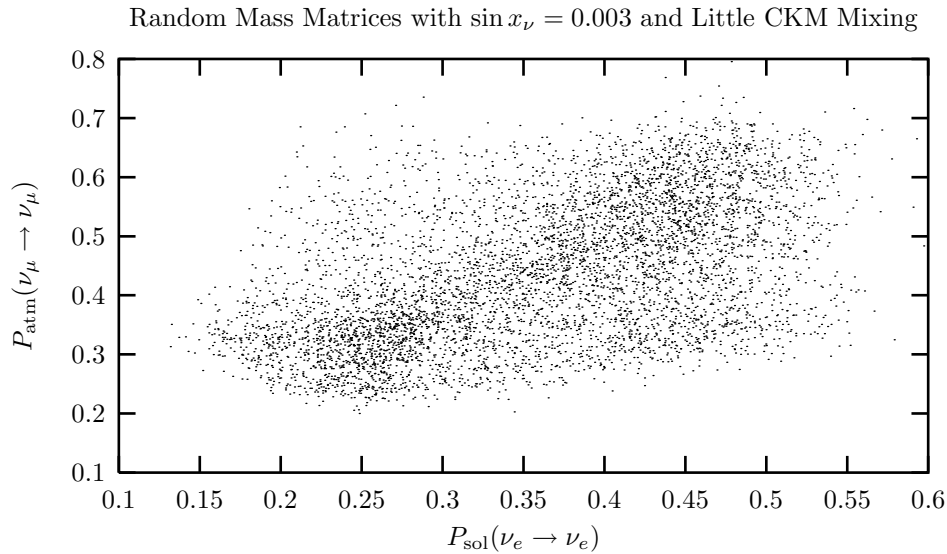}
\caption{The probabilities \( P_{\mathrm{sol}}(\nu_e \to \nu_e ) \)
and \( P_{\mathrm{atm}}(\nu_\mu \to \nu_\mu ) \)
for 10000 random mass matrices \(\cM\)
all with the parameter \(\sin x_\nu = 0.003\)
and with inter-generational mixing suppressed
by factors of 0.2.}
\label {fig:c4}
\end{figure}
\begin{figure}
\centering
\includegraphics{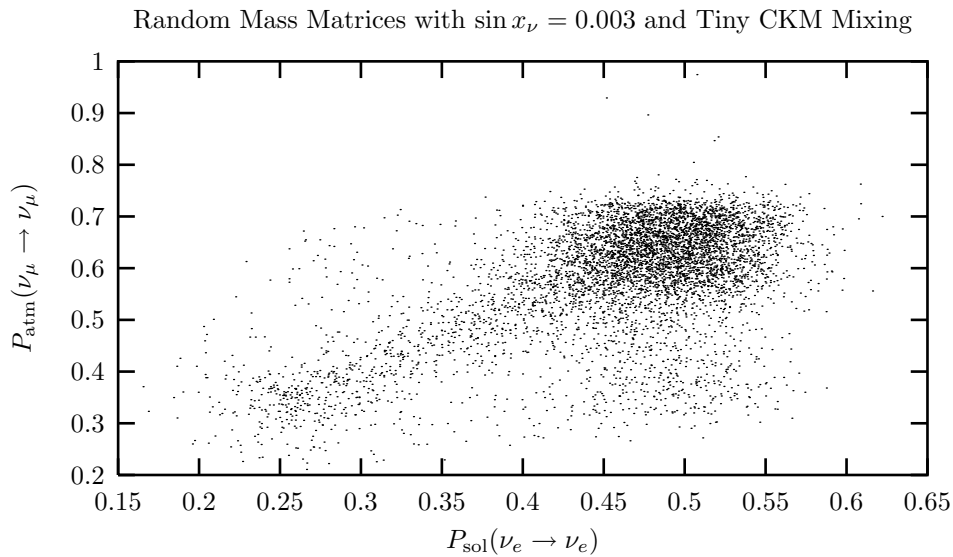}
\caption{The probabilities \( P_{\mathrm{sol}}(\nu_e \to \nu_e ) \)
and \( P_{\mathrm{atm}}(\nu_\mu \to \nu_\mu ) \)
for 10000 random mass matrices \(\cM\)
all with the parameter \(\sin x_\nu = 0.003\)
and with inter-generational mixing suppressed
by factors of 0.05.}
\label{fig:c5}
\end{figure}

The probabilities \( P_{\mathrm{sol}}(\nu_e \to \nu_e ) \) 
and \( P_{\mathrm{atm}}(\nu_\mu \to \nu_\mu ) \)
tend to be somewhat larger when inter-generational mixing is limited.  
In Fig.~\ref{fig:c4} these probabilities are displayed for
10000 random mass matrices with \( \sin x_\nu = 0.003 \)
and with the singly off-diagonal matrix elements of \(D, E,\) and \(F\) 
suppressed by 0.2 and the doubly off-diagonal matrix elements
suppressed by 0.004\@.
The points in Fig.~\ref{fig:c4} are 
in much better agreement with (\ref{bothexp})
than are those of Fig.~\ref{fig:e4},
but they still fail to match the experiments.
Yet if the suppression of mixing is more severe,
with factors respectively of 0.05 and 
0.0025, then as shown in Fig.~\ref{fig:c5}
the probabilities \( P_{\mathrm{sol}}(\nu_e \to \nu_e ) \)
and \( P_{\mathrm{atm}}(\nu_\mu \to \nu_\mu ) \)
do tend to cluster around \((\half,\frac{2}{3})\) as required
by the data.
\par
\begin{figure}
\centering
\includegraphics{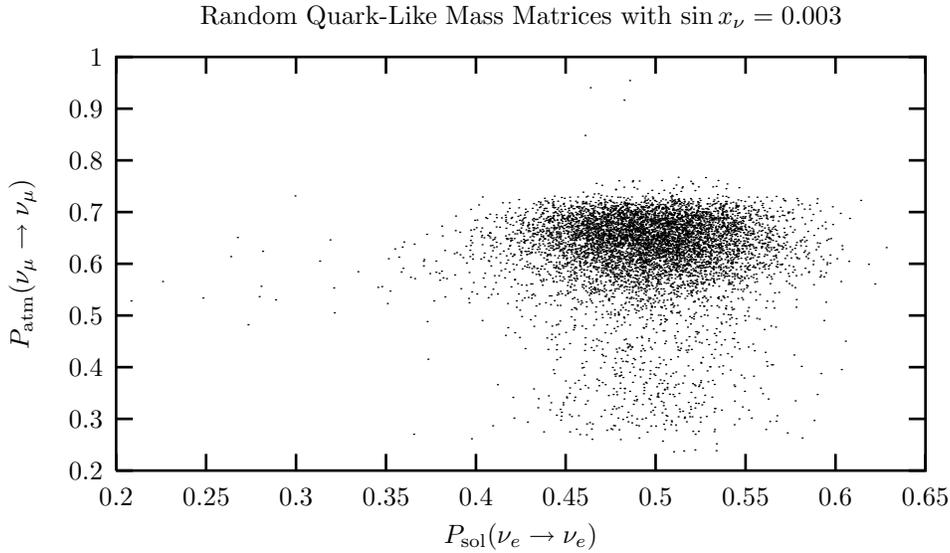}
\caption{The probabilities \( P_{\mathrm{sol}}(\nu_e \to \nu_e ) \)
and \( P_{\mathrm{atm}}(\nu_\mu \to \nu_\mu ) \)
for 10000 random mass matrices \(\cM\)
all with the parameter \(\sin x_\nu = 0.003\),
with inter-generational mixing suppressed
by factors of 0.2, and
with a quark-like mass hierarchy.}
\label {fig:h5}
\end{figure}

It is possible to relax the factors that suppress
inter-generational mixing back to 0.2 and 0.04
and improve the agreement with the
experimental constraints (\ref {solexp}) and (\ref {atmexp})
(while satisfying the CHOOZ constraint)
provided that one also requires 
that there be a quark-like mass hierarchy.
The points in Fig.~\ref{fig:h5}
were generated from 10000 random mass matrices
\( \cM \) with \( \sin x_\nu = 0.003 \)
and CKM-suppression factors
of 0.2 and 0.04 as in Fig.~\ref{fig:c4} 
and with the \(i,j\)-th elements
of the mass matrices \( E, F,\) and \(D\) scaled
by the factors \(f(i)*f(j)\) where
\( \vec f = ( 0.2, 1, 2 ) \)\@.
Thus the mass matrix \(\cM\)
has the \( \tau, \tau \)
elements that are larger than its
\( \mu, \mu \) elements and \( \mu, \mu \) elements
that in turn are larger than its \( e, e \) elements. 
The clustering  of 
the probabilities \( P_{\mathrm{sol}}(\nu_e \to \nu_e ) \)
and \( P_{\mathrm{atm}}(\nu_\mu \to \nu_\mu ) \) 
around \((\half,\frac{2}{3})\) in Fig.~\ref{fig:h5} shows that 
the experimental results (\ref {solexp}) and (\ref {atmexp}) 
are satisfied.
The vector \(\vec f\)
was tuned so as to nearly saturate 
the 
cosmological upper bound (\ref {eubound}) of about \( 8 \, \mathrm{eV}\)\@.
\subsection*{LSND, KARMEN2, and MiniBooNE}
\par
\begin{figure}
\centering
\includegraphics{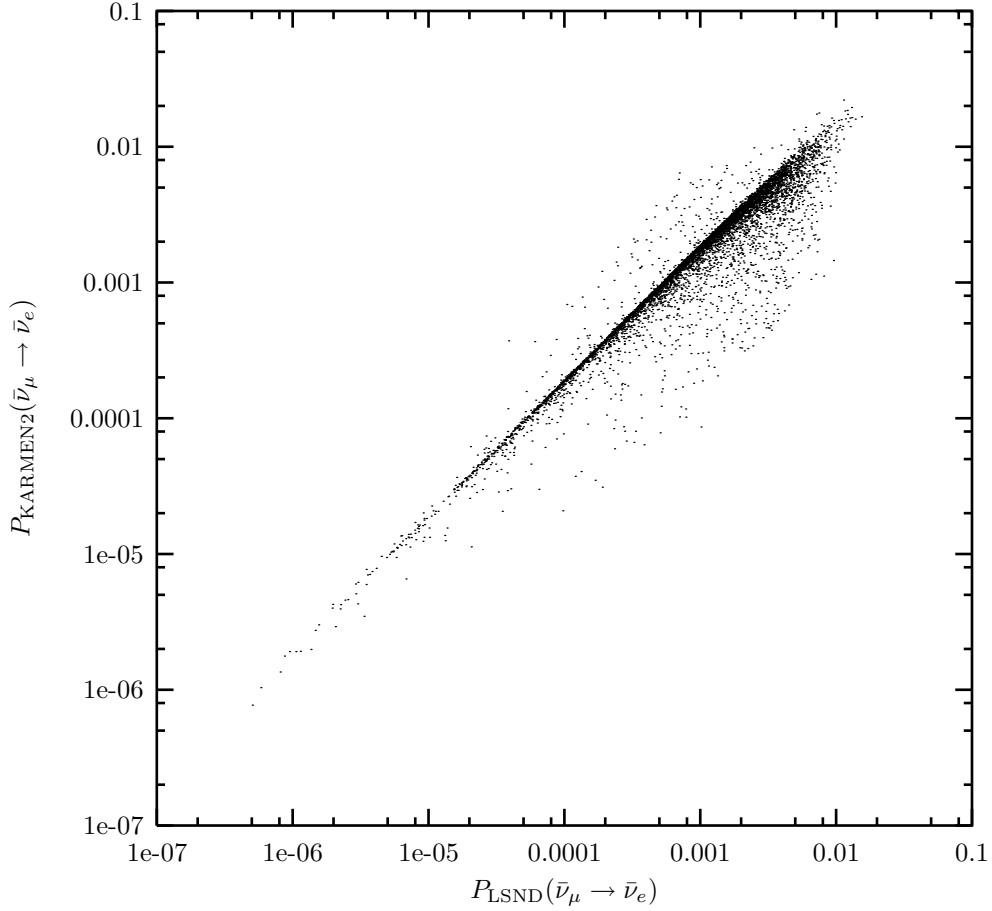}
\caption{The probabilities \( P_{\mathrm{LSND}}(\bar\nu_\mu \to \bar\nu_e ) \)
and \( P_{\mathrm{KARMEN2}}(\bar\nu_\mu \to \bar\nu_e ) \)
for 10000 random mass matrices \(\cM\)
all with the parameter \(\sin x_\nu = 0.003\),
with inter-generational mixing suppressed
by factors of 0.2, and
with a quark-like mass hierarchy.}
\label {fig:ls}
\end{figure}

Because \( x_\nu\) is small,
and because neutrinos oscillate mainly
into sterile neutrinos,
the probabilities of the appearance of neutrinos
are small, as shown by LSND and by KARMEN2\@.  
But since the mass differences among
the three nearly degenerate pairs of neutrinos
can lie in the eV range, such oscillations should
be observable.
In Fig.~\ref{fig:ls}
for a set of 10000 mass matrices generated randomly
with the same parameters as for Fig.~\ref{fig:h5},
the probabilities \(P(\bar\nu_\mu \to \bar\nu_e)\)
are shown for LSND and KARMEN2.
\par
\begin{figure}
\centering
\includegraphics{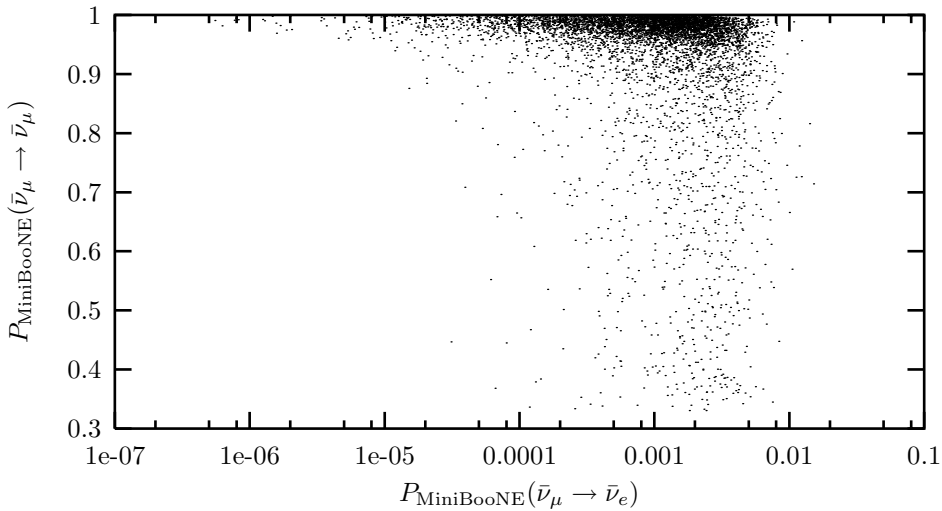}
\caption{The probabilities 
\( P_{\mathrm{MiniBooNE}}(\bar\nu_\mu \to \bar\nu_e ) \)
and \( P_{\mathrm{MiniBooNE}}(\bar\nu_\mu \to \bar\nu_\mu ) \)
for 10000 random mass matrices \(\cM\)
all with the parameter \(\sin x_\nu = 0.003\),
with inter-generational mixing suppressed
by factors of 0.2, and
with a quark-like mass hierarchy.}
\label {fig:bo}
\end{figure}

In Fig.~\ref{fig:bo}
for a set of 10000 mass matrices generated randomly
with the same parameters as for Figs.~\ref{fig:h5}
and \ref{fig:ls},
the probabilities \(P(\bar\nu_\mu \to \bar\nu_e)\)
and \(P(\bar\nu_\mu \to \bar\nu_\mu)\) for
MiniBooNE are plotted.
If intergenerational mixing
is suppressed only by factors of 0.2
and if MiniBooNE can achieve a sensitivity
of 0.001 for \(\bar\nu_\mu \to \bar\nu_e\)
and a precision of 0.01 for \(\bar\nu_\mu \to \bar\nu_\mu\),
then it has a good chance of seeing
both the appearance of \(\bar\nu_e\)
and the disappearance of \(\bar\nu_\mu\)\@.
\begin{figure}
\centering
\includegraphics{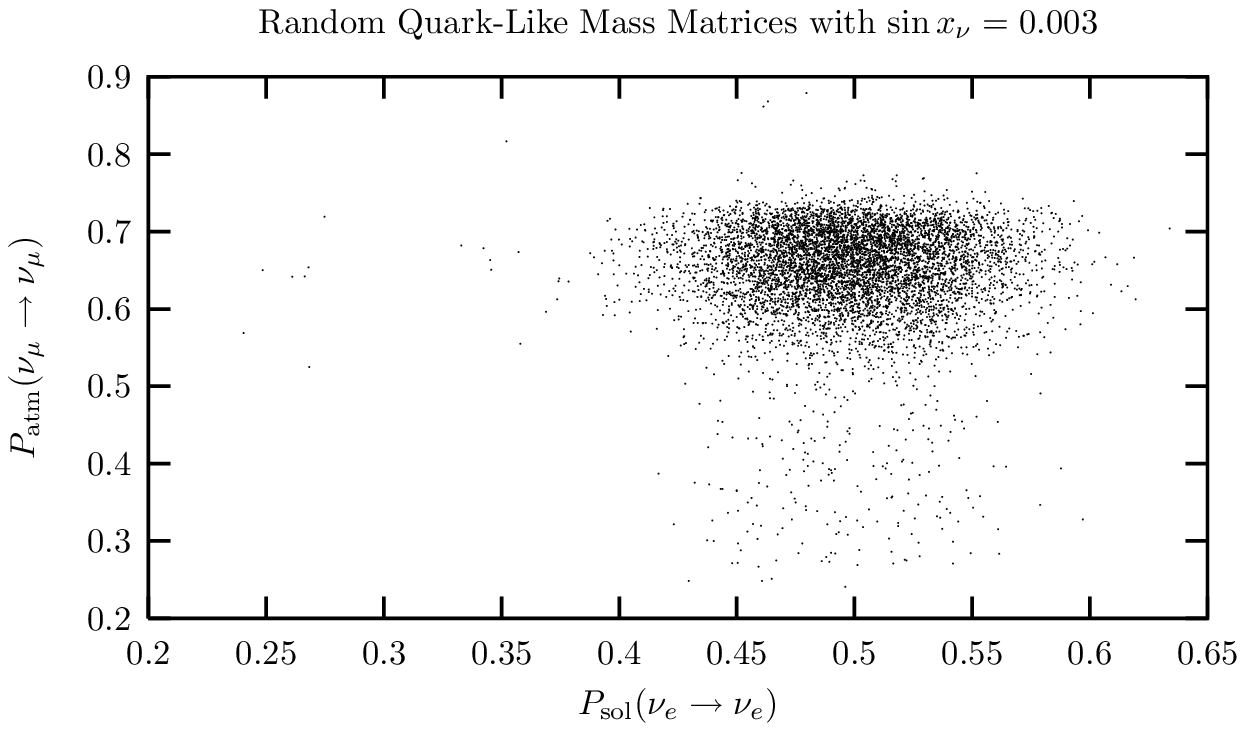}
\caption{The probabilities \( P_{\mathrm{sol}}(\nu_e \to \nu_e ) \)
and \( P_{\mathrm{atm}}(\nu_\mu \to \nu_\mu ) \)
for 10000 random mass matrices \(\cM\)
all with the parameter \(\sin x_\nu = 0.003\),
with inter-generational mixing suppressed
by factors of 0.1, and
with a quark-like mass hierarchy.}
\label {fig:h6}
\end{figure}
\begin{figure}
\centering
\includegraphics{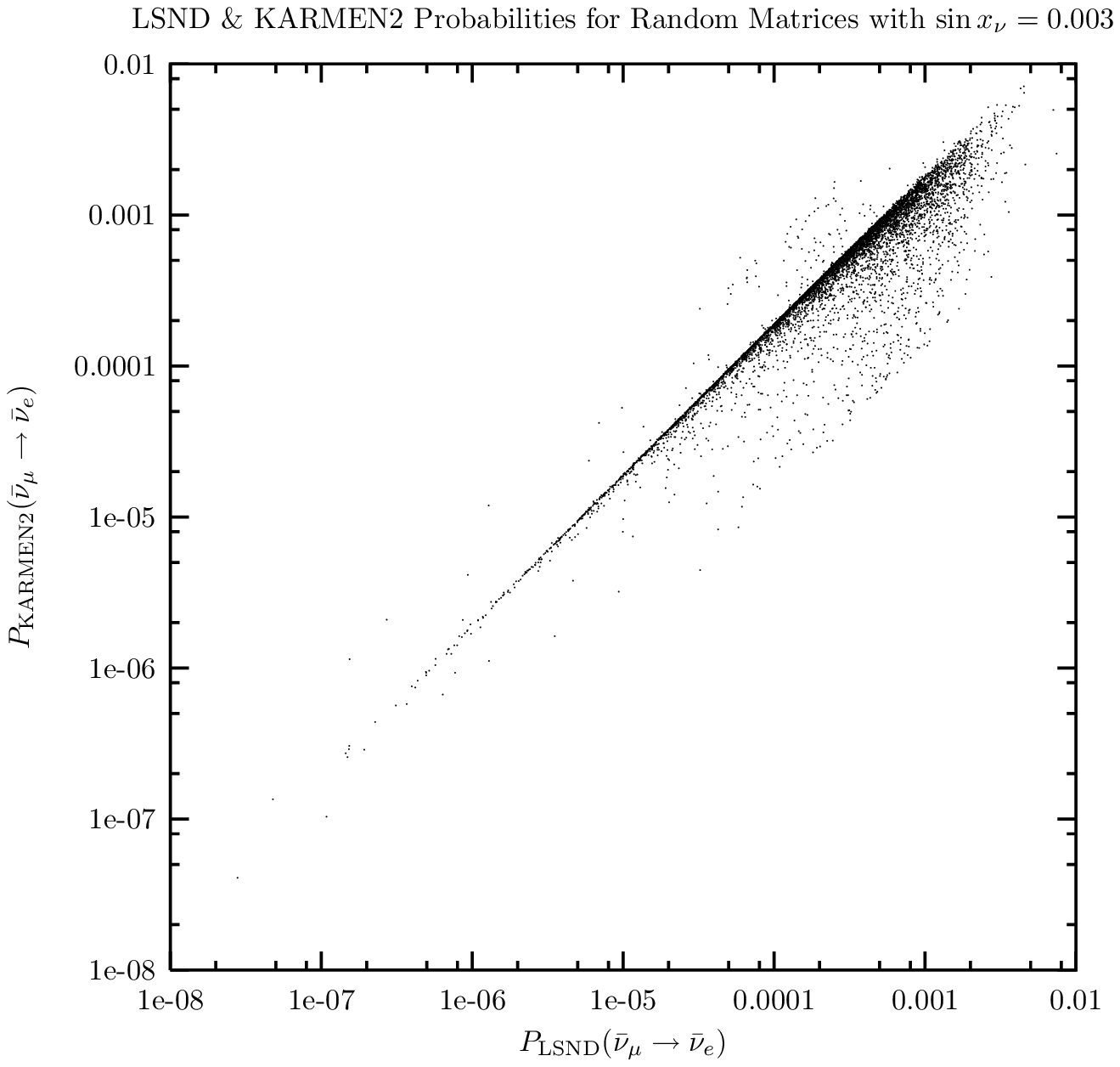}
\caption{The probabilities \( P_{\mathrm{LSND}}(\bar\nu_\mu \to \bar\nu_e ) \)
and \( P_{\mathrm{KARMEN2}}(\bar\nu_\mu \to \bar\nu_e ) \)
for 10000 random mass matrices \(\cM\)
all with the parameter \(\sin x_\nu = 0.003\),
with inter-generational mixing suppressed
by factors of 0.1, and
with a quark-like mass hierarchy.}
\label {fig:l2}
\end{figure}
\begin{figure}
\centering
\includegraphics{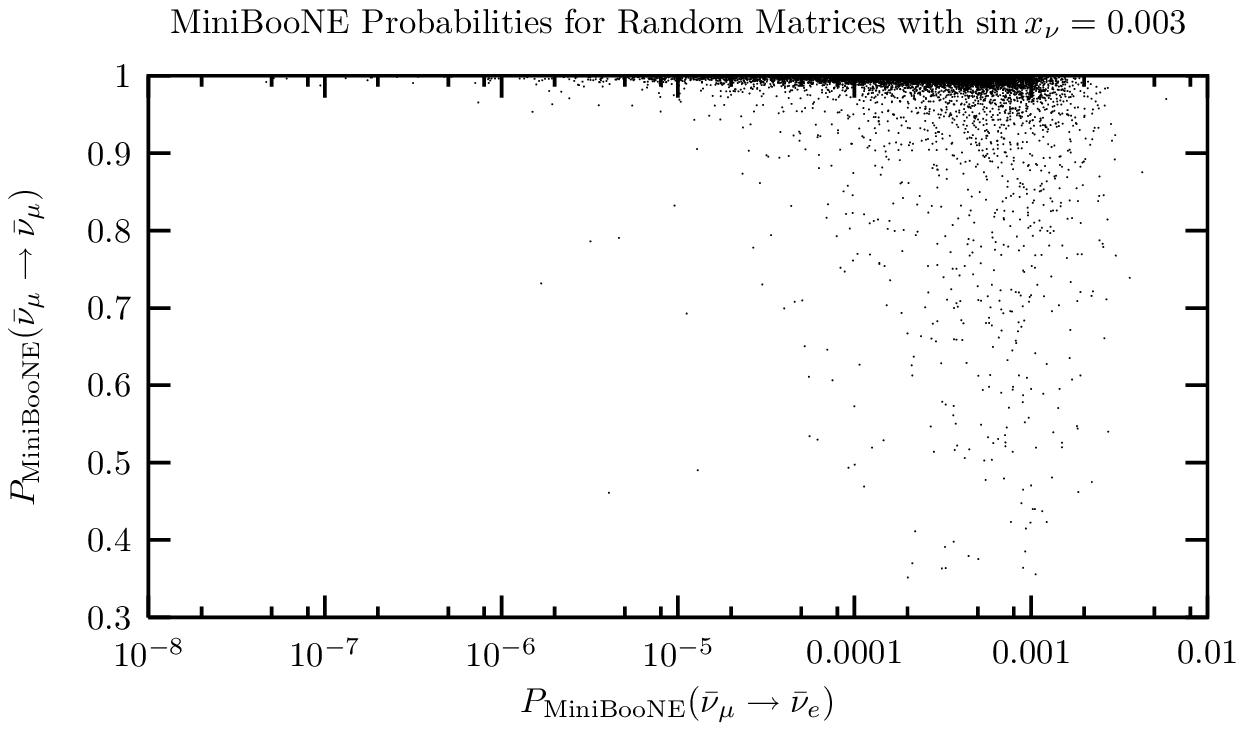}
\caption{The probabilities 
\( P_{\mathrm{MiniBooNE}}(\bar\nu_\mu \to \bar\nu_e ) \)
and \( P_{\mathrm{MiniBooNE}}(\bar\nu_\mu \to \bar\nu_\mu ) \)
for 10000 random mass matrices \(\cM\)
all with the parameter \(\sin x_\nu = 0.003\),
with inter-generational mixing suppressed
by factors of 0.1, and
with a quark-like mass hierarchy.}
\label {fig:bo2}
\end{figure}
\par
It may be, however, that
intergenerational mixing 
is suppressed only by factors of 0.1.
For as shown in Fig.~\ref{fig:h6},
a set of 10000 matrices randomly generated,
with CKM suppression factors of 0.1 for the
off-diagonal terms and 0.01 for the off-off-diagonal ones
and with the mass-hierarchy vector \(\vec f = (.2,1,2)\),
agrees with the solar and atmospheric experiments
just as well as those of Fig.~\ref{fig:h5}.
With these parameters the probabilities
\(P(\bar\nu_\mu \to \bar\nu_e ) \) at
LSND and KARMEN2 are smaller,
as shown in Fig.~\ref{fig:l2}\@.
The chances of detection at MiniBooNE
are similarly reduced as shown in Fig.~\ref{fig:bo2}\@.

\subsection*{Neutrinoless Double Beta Decay}
\par
Neutrinoless double beta decay occurs when
a right-handed antineutrino emitted in one
decay \( n \to p + e^- + \bar \nu_e \)
is absorbed as a left-handed neutrino in 
another decay \( \nu_e + n \to p + e^- \)\@. 
To lowest order these decays proceed via
the Majorana mass term 
\( - i F_{ee}^* \nu_e^\dgr \s_2 \nu_e^\dgr \)\@.
Let us introduce a second angle \( y_\nu\)
defined by
\beq
\sin^2 y_\nu = \frac{\mathrm{Tr}( F^\dgr F )}
{\mathrm{Tr}( E^\dgr E + F^\dgr F )}.
\label {sf}
\eeq
We have seen that we may fit the experimental data
(\ref{solexp}) and (\ref{atmexp})
by assuming that \(\sin x_\nu \simeq 0.003\)
and by requiring the mass matrices 
\( E, F, \) and \(D\) to exhibit
quark-like mass hierarchies with little inter-generational mixing.
Under these conditions
the rate of \(0\nu\b\b\) decay is
limited by the factor
\beq
|F_{ee}|^2 \lsim \sin^2 x_\nu \, \sin^2 y_\nu \, m_{\nu_e}^2,
\label {0vbb}
\eeq
in which \( m_{\nu_e} \) is the heavier of the 
lightest two neutrino masses.
Thus the rate of \( 0\nu \b\b \) decay 
is suppressed by an extra factor 
\(\sim \sin^2 x_\nu \sin^2 y_\nu \lsim 10^{-5}\)
resulting in lifetimes \(T_{\half,0\nu\b\b} > 2\times10^{27}\,\)yr\@.
The \(B-L\) model therefore explains why
neutrinoless double beta decay has not been seen
and predicts that the current and upcoming experiments
Heidelberg/Moscow, IGEX, GENIUS, and CUORE
will not see \( 0\nu \b\b \) decay.

\section{Conclusions}
The standard model slightly extended to include
right-handed neutrino fields exactly conserves \(B-L\)
if all Majorana mass terms vanish.
It is therefore natural~\cite{tHooft} to assume that the Majorana mass terms 
are small compared to the Dirac mass terms.
An angle \( x_\nu \)
is introduced that characterizes the relative
importance of these two kinds of mass terms.
When this parameter is very small, then the neutrinos
are nearly Dirac and only slightly Majorana.
In this case the six neutrino masses \(m_j\)
coalesce into three pairs of nearly degenerate masses.
Thus the very tiny mass differences seen in the
solar and atmospheric neutrino experiments
are simply explained by the natural assumption that
\( x_\nu \approx 0 \) or equivalently that \(B-L\) 
is almost conserved.
\par
In these experiments
the probabilities \( P_{\mathrm{sol}}(\nu_e \to \nu_e ) \)
and \( P_{\mathrm{atm}}(\nu_\mu \to \nu_\mu ) \) are respectively
approximately one half and two thirds.
One may fit these probabilities with random mass matrices 
in the eV range by setting \( \sin x_\nu = 0.003 \)
and requiring the neutrino mass matrices
\( E, F, \) and \(D\) to exhibit
quark-like mass hierarchies with little inter-generational mixing.
\par
The three mass differences among the three 
nearly degenerate pairs are constrained only
by the cosmological bound \( \sum_j m_j \lsim 8 \,\)eV
and may lie in the range \([0.01,4]\,\)eV\@. 
\par
This \(B-L\) model leads to these predictions: 
\begin{enumerate}
\item Because \( \sin^2 x_\nu \approx 0 \) and
because inter-generational mixing is suppressed,
neutrinos oscillate mainly into sterile neutrinos 
of the same flavor and not into neutrinos of other flavors. 
Hence rates for the appearance of neutrinos, \(P(\nu_i \to \nu_{i'})\)
with \( i \ne i' \), are very low,
as shown by LSND and by KARMEN2\@.  
But because the mass differences among the three nearly
degenerate pairs of masses can lie in the eV range,
MiniBooNE may detect \(\bar\nu_\mu \to \bar\nu_e\)\@.
\item The assumption that \( \sin^2 x_\nu \)
is very small naturally explains the very small differences
of squared masses seen in the solar and atmospheric experiments
without requiring that the neutrino masses themselves be very small.
Thus the neutrinos may very well saturate 
the cosmological bound, \( \sum_j m_j \lsim 8 \, \mathrm{eV} \)\@.
In fact the masses associated with the points of 
Figs.~\ref{fig:h5}--\ref{fig:bo2}
do nearly saturate this bound.  Neutrinos thus may well be
an important part of hot dark matter.
\item The disappearance of \( \nu_\tau \) should \emph{in principle}
be observable.
\item In the \(B-L\) model, the rate of neutrinoless
double beta decay
is suppressed by an extra factor 
\(\sim \sin^2 x_\nu \, \sin^2 y_\nu \lsim 10^{-5}\)
resulting in lifetimes greater than \(2\times10^{27}\,\)yr.
Thus the current and upcoming experiments
Heidelberg/Moscow, IGEX, GENIUS, and CUORE
will not see \( 0\nu \b\b \) decay.
\end{enumerate}

\section*{Acknowledgements}
I am grateful to H.~Georgi for
a discussion of neutrinoless double beta decay;
to Byron Dieterle, Klaus Eitel, and Randolph Reeder
for information about LSND and KARMEN2;
to Naoshi Sugiyama for a discussion of cosmological bounds
on neutrino masses;
to Thomas Gaisser and Todor Stane for information
about the Bartol fluxes;
to Mark Messier for details about Super-Kamiokande;
and to Bernd Bassalleck, Christy Crowley, James Demmel, 
Michael Gold, Gary Herling, Dean Karlen,
Boris Kayser, Plamen Krastev, Don Lichtenberg, Steven McCready,
Rabindra Mohapatra, Steven Riley, Dmitri Sergatskov,
and Gerard Stephenson
for other helpful conversations.
Some of the computations of this paper were performed
on the Black Bear Linux Cluster
of the Albuquerque High-Performance Computing Center
of the University of New Mexico.

\appendix
\section*{Appendix: LAPACK}
The thousands of \(6\times6\)
matrix computations displayed in the several graphs
were performed 
in the inexpensive Lintel computing environment of
Intel chips, Red Hat Linux, the Portland Group's Fortran compiler, 
and the linear-algebra software of the LAPACK collaboration.
This appendix describes how to use 
LAPACK to perform the singular-value decomposition of an arbitrary 
6x6 complex matrix
and the Takagi factorization of a 6x6 complex symmetric matrix.
\par
The driver subroutine ZGESVD~\cite{lapack} does a singular-value
decomposition of an arbitrary matrix A.
The FORTRAN 90 call is
\begin{quote}
call ZGESVD( JOBU, JOBVT, M, N, A, LDA, S, U, LDU, \&
\hspace{1.5in}   VT, LDVT, WORK, LWORK, RWORK, INFO )
\end{quote}
in which JOBU = 'A', JOBVT = 'A', M = N = LDA = LDVT = 6,
WORK is an LWORK-dimensional double-complex vector, and
RWORK is a 26-dimensional double-precision real vector.
ZGESVD performs a singular-value decomposition
of the \(6\times6\) double-complex matrix A
which is the  mass matrix \(\cM\) and reports
its singular values, which are the neutrino masses \(m_j\),
as the 6-dimensional double-precision real vector S.
The matrices \(U\) and \(V^\dgr\) are contained
in the \(6\times6\) double-complex matrices U and VT.
The integer INFO describes the level of success
of the computation.
The best value of the system-dependent integer LWORK is
reported as the real part of WORK(1);
for the computations of this paper, which were done
on Pentium II's and III's, LWORK was 396\@.
If JOBU is 'A' and JOBVT is 'A', then the subroutine call
destroys the matrix A.  Thus one must redefine A before
every call to ZGESVD.
\par
If the matrix A is symmetric, then one may
convert the singular-value decomposition
\(A = U S V^\dgr\)
into the Takagi factorization \(A = W S W^\top\)
by the FORTRAN 90 equivalent of
\beq
W_{ij} = U_{ij} \, \sqrt{\frac{V^\dgr_{jk}}{|V^\dgr_{jk}|}
\,\frac{|U_{kj}|}{U_{kj}}}
\label {SVDtoTF}
\eeq
where the index \(k\) is chosen to avoid
any possible singularity.


\begin{thebibliography}{99}
\bibitem{mgmrm} M.~Gell-Mann, P.~Ramond, and R.~Slansky,
in {\sl Supergravity\/} (Proceedings of the Supergravity
Workshop at Stony Brook, 1979, ed. by P.~van Nieuwenhuisen and 
D.~Z. Freedman, North-Holland, Amsterdam, 1979), p.~315,
and R.~N. Mohapatra and G.~Senjanovic, 
{\sl Phys.~Rev.~Lett.\/} 44, 912 (1980).
\bibitem{Wolfenstein} L.~Wolfenstein, {\sl Nucl.~Phys.\/} B186 (1981) 147;
S.~M. Bilenky and B.~M. Pontecorvo, 
{\sl Yad.~Fiz.\/} 38 (1983) 415
({\sl Sov.~J.~Nucl.~Phys.\/} 38 (1983) 248);
S.~M. Bilenky and S.~T. Petcov,
{\sl Rev.~Mod.~Phys.\/} 59 (1987) 671.
\bibitem{PDauthors} 
C.~Giunti, C.~W. Kim, and U.~W. Lee,
{\sl Phys.~Rev.\/} D46 (1992) 3034;
J.~P. Bowes and R.~R. Volkas,
hep-ph/9804310 =
{\sl J.~Phys.\/} G24 (1998) 1249;
R.~Foot and R.~R. Volkas,
hep-ph/9505359 =
{\sl Phys.~Rev.\/} D52 (1995) 6595.
\bibitem{Geiser} A.~Geiser, hep-ph/9901433
= {\sl Phys.~Lett.\/} B444 (1998) 358.
\bibitem{Horn} R.~A. Horn and C.~A. Johnson,
{\sl Matrix Analysis\/} (Cambridge University Press, Cambridge, 1985):
p.~411.
\bibitem{HornJohnson} R.~A. Horn and C.~A. Johnson,
\emph{op. cit.}: p.~204. 
\bibitem{Takagi} T.~Takagi, {\sl Japan. J.~Math.\/} 1 (1925) 83.
\bibitem{tHooft} G.~'t Hooft, in {\sl Recent Developments in Gauge Theories\/}
(Proceedings of the 1979 Carg\`{e}se Summer Institute,
Carg\`{e}se, France, ed. by G.~'t Hooft {\sl et al.,}
NATO Advanced Study Institute Series B: Physics Vol. 59
(Plenum Press, New York, 1980), p.~135\@.
See also H.~Georgi, {\sl Weak Interactions
and Modern Particle Theory\/}
(Benjamin/Cummings, 1984), p.~127.
\bibitem{LAPACK} E.~Anderson, Z.~Bai, C.~Bischof, S.~Blackford,
J.~Demmel, J.~Dongarra, J.~Du Croz, A.~Greenbaum, S.~Hammarling,
A.~McKenney, and D.~Sorensen, 
{\sl LAPACK Users' Guide\/} 
(3d ed., SIAM, Philadelphia, PA, 1999) available online at
www.netlib.org/lapack/lug/lapack\_lug.html.
\bibitem{lapack} The subroutine ZGESVD and all the 
subroutines and functions on which it depends are freely
available from www.netlib.org/lapack/
and come prebuilt with the Portland Group's FORTRAN90
compiler which may be downloaded
from www.pgroup.com/.
\bibitem{SPQR} S.~P. Rosen, {\sl Phys.~Rev.\/}D29 (1984) 2535.
\bibitem{SS} G.~D. Starkman and D.~Stojkovic, hep-ph/9909350.
\bibitem{SWII}
S.~Weinberg, {\sl The Quantum Theory of Fields\/},
vol. II (Cambridge University Press, 1996), p.~308.
\bibitem{LEP}
The LEP Collaboration and the
LEP Electroweak Working Group, as reported by J.~Mnich
at the International Europhysics Conference, 
Tampere, Finland (July 1999).
\bibitem{FKM} P.~Fisher, B.~Kayser, and K.~S. McFarland,
hep-ph/9906244 and {\sl Ann.~Rev.~Nucl.~Part.~Sci.\/}
49 (1999, in press).
\bibitem{Kolb} E.~W. Kolb and M.~S. Turner,
{\sl The Early Universe\/}
(Addison-Wesley, Redwood City, CA, 1990), p.~123.
\bibitem{FLS} M.~Fukugita, G.-C.~Liu, and N.~Sugiyama,
{\sl Phys.\ Rev.\ Letters\/} 84 (2000) 1082.
\bibitem{SDSS} SDSS: www.astro.princeton.edu/PBOOK/welcome.htm
\bibitem{HET} W.~Hu, D.~J. Eisenstein, \& M.~Tegmark,
{\sl Phys.\ Rev.\ Letters\/} 80 (1998) 5255.
\bibitem{BBNlimit} A.~Dolgov, {\sl Sov.\ J.\ Nucl.\ Phys.\/} 33, 700 (1981);
R.~Barbieri and A.~Dolgov, {\sl Phys.\ Lett.\/} B237, 440 (1990);
{\sl Nucl.\ Phys.\/} B349, 743 (1991);
K.~Enqvist, K.~Kainulainen and M.~Thomson, 
{\sl Nucl.\ Phys.\/} B373, 498 (1992);
J.~Cline, {\sl Phys.\ Rev.\ Lett.\/} 68, 3137 (1992);
X.~Shi, D.~N. Schramm and B.~D. Fields, 
{\sl Phys.\ Rev.\/} D48, 2568 (1993).
\bibitem{FootVolkas} R.~Foot \& R.~R. Volkas,
{\sl Phys.\ Rev.\ Letters\/} 75 (1995) 4350;
R.~Foot, M.~J. Thomson, \& R.~R. Volkas,
{\sl Phys.\ Rev.\/} D53 (1996) R5349;
R.~Foot \& R.~R. Volkas,
{\sl Phys.\ Rev.\/} D55 (1997) 5147;
D56 (1997) 6653.
\bibitem{MaxMatias} M.~Tegmark \& M.~Zaldarriaga,
astro-ph/0004393. 
\bibitem{Wright} K.~Dick, M.~Lindner, M.~Ratz, and D.~Wright,
{\sl Phys.\ Rev.\ Letters\/} 84 (2000) 4039.
\bibitem{sphalerons} Ref.\cite{SWII}, pp.~455 and 476.
\bibitem{AD} I.~A. Affleck \& M.~Dine,
{\sl Nucl.\ Phys.\/} B249 (1985) 361.
\bibitem{CCG} A.~Casas, W.~Y. Cheng, \& G.~Gelmini,
{\sl Nucl.\ Phys.\/} B538 (1999) 297.
\bibitem{MD} J.~McDonald,
{\sl Phys.\ Rev.\ Letters\/} 84 (2000) 4798. 
\bibitem{Cabibbo} N.~Cabibbo, {\sl Phys.\ Lett.\/} 72B (1978) 333.  
\bibitem{Boris} B.~Kayser, in C.~Caso \emph{et al.},
{\sl Eur.~Phys.~J.\/} C3 (1998) 1.
\bibitem{Gaisser} V.~Agrawal, T.~K. Gaisser, P.~Lipari,
and T.~Stanev, {\sl Phys.~Rev.\/} 53 (1996) 1314.
\end{thebibliography}
\end{document}